\documentclass[final,5p,times,twocolumn]{elsarticle}

\usepackage{amssymb}
\usepackage{amsmath}
\usepackage{wasysym}
\usepackage{url}
\usepackage{todonotes}

\DeclareMathOperator\erf{erf}

\newcounter{bla}

\journal{Computer Physics Communications}

\begin{document}

\begin{frontmatter}

  \title{A geometric multigrid library for quadtree/octree AMR grids coupled to MPI-AMRVAC}

  \author[a,b]{J.~Teunissen\corref{author}}
  \author[a]{R.~Keppens}

  \cortext[author] {Corresponding author.\\\textit{E-mail address:}
    jannis@teunissen.net}

  \address[a]{Centre for mathematical Plasma Astrophysics, Department of
    Mathematics, KU Leuven, Celestijnenlaan 200B, 3001 Leuven, Belgium}

  \address[b]{Centrum Wiskunde \& Informatica, PO Box 94079, 1090 GB Amsterdam,
    The Netherlands}

  \begin{abstract}
    We present an efficient MPI-parallel geometric multigrid library for
    quadtree (2D) or octree (3D) grids with adaptive refinement. Cartesian 2D/3D
    and cylindrical 2D geometries are supported, with second-order
    discretizations for the elliptic operators. Periodic, Dirichlet, and Neumann
    boundary conditions can be handled, as well as free-space boundary
    conditions for 3D Poisson problems, for which we use an FFT-based solver on
    the coarse grid. Scaling results up to 1792 cores are presented. The library
    can be used to extend adaptive mesh refinement frameworks with an elliptic
    solver, which we demonstrate by coupling it to \texttt{MPI-AMRVAC}. Several
    test cases are presented in which the multigrid routines are used to control
    the divergence of the magnetic field in magnetohydrodynamic simulations.
  \end{abstract}


\end{frontmatter}

\section{Introduction}

A typical example of an elliptic partial differential equation (PDE) is
Poisson's equation
\begin{equation}
  \label{eq:poisson}
  \nabla \cdot (\varepsilon \nabla \phi) = f,
\end{equation}
where the right-hand side $f$ and coefficient $\varepsilon$ are given and $\phi$
has to be obtained given certain boundary conditions. Equations like
(\ref{eq:poisson}) appear in many applications, for example when computing
electrostatic or gravitational potentials, or when simulating incompressible
flows. An important property of elliptic equations is that they are non-local:
their solution at one location depends on the solution and right-hand side
elsewhere. Here we present a library for the parallel solution of elliptic PDEs
on quadtree and octree grids with adaptive mesh refinement (AMR).

Elliptic PDEs can be solved with e.g., fast Fourier transforms (FFTs), cyclic
reduction methods, direct sparse solvers, preconditioned iterative methods,
multipole methods and multigrid methods, see e.g.~\cite{Gholami_2016}. These
methods differ in their flexibility, for example in terms of supported mesh
types and boundary conditions, and in whether the coefficient $\varepsilon$ in
equation (\ref{eq:poisson}) is allowed to have a smooth or discontinuous spatial
variation. They also differ in their efficiency. The fastest multigrid methods
operate in time $O(N)$, where $N$ denotes the number of unknowns. FFT-based
methods (with or without cyclic reduction) typically require time $O(N \log N)$.
Most other methods are more expensive, although their cost is often
problem-dependent. Due to the non-local nature of elliptic equations, there are
also significant differences in how well solvers can be parallelized. A
comparison of the performance and scaling of several state-of-the-art Poisson
solvers can be found in~\cite{Gholami_2016}.


Our motivation was to extend \texttt{MPI-AMRVAC}~\cite{Xia_2018,porth_2014} with an
elliptic solver. \texttt{MPI-AMRVAC} is a parallel AMR framework for
(magneto)hydrodynamics simulations that it is typically used to study solar and
astrophysical phenomena. The framework has a focus on solving conservation
laws with shock-capturing methods and quadtree/octree AMR. Such AMR grids
are ideally suited to geometric multigrid methods, which iteratively solve
elliptic equations by employing a hierarchy of grids. Geometric multigrid
methods can also be highly efficient, with an ideal $O(N)$ time complexity, and
they are matrix-free, which means that no matrix has to be stored or
pre-computed.

There are already a number of AMR frameworks that include a multigrid solver.
Examples are \texttt{Boxlib}~\cite{Zhang_2016} (superseded by \texttt{AMReX}),
\texttt{Maestro}~\cite{Almgren_2009}, \texttt{Gerris}~\cite{Popinet_2003},
\texttt{RAMSES}~\cite{Teyssier_2002}, \texttt{NIRVANA}~\cite{Ziegler_2008} and
\texttt{Paramesh}/\texttt{FLASH}~\cite{Fryxell_2000,ricker_2008}. However, the
included multigrid solvers are typically coupled to (and optimized for) the
application codes, so that they cannot easily be used in other projects.

In recent years, several highly scalable multigrid solvers have been developed.
A combination of geometric and algebraic multigrid was used
in~\cite{Sundar_2012} to obtain a matrix-free method that could scale to
$2.6 \times 10^5$ cores. Relevant is also the development of the
open-source \texttt{HPGMG} code~\cite{Adams_2014} (\url{https://hpgmg.org/}), which is aimed
at benchmarking HPC systems with geometric multigrid methods. \texttt{HPGMG} has
already been coupled to \texttt{Boxlib}, but as the code's primary goal appears
to be benchmarking it was not clear how easily it could be coupled to
\texttt{MPI-AMRVAC}, which is written in Fortran. Another relevant code is
\texttt{DENDRO}~\cite{sampath_2008}, which can solve PDEs on finite element
meshes. \texttt{DENDRO} was written in \texttt{C++} and uses the \texttt{PETSc}
library~\cite{petsc-web-page}. Finally, we mention \texttt{Hypre}~\cite{falgout:hypre}, a
library of high performance multigrid solvers and preconditioners.

Because there appeared to be no geometric multigrid library that we could easily
couple to \texttt{MPI-AMRVAC}, we have developed such a library ourselves. The
main features of the library are:
\begin{itemize}
  \item Support for solving elliptic PDEs on quadtree/octree AMR grids in
  Cartesian (2D/3D) and axisymmetric (2D) geometries.
  \item Support for Dirichlet, Neumann and periodic boundary conditions, as well
  as free space boundary conditions in 3D.
  \item MPI-based parallelization that can scale to $10^3$ or more processors.
  \item All source code is written in Fortran, under an open source license
  (GPLv3). The source code can be found at \url{https://github.com/jannisteunissen/octree-mg}.
\end{itemize}
The library is relatively simple and small, with currently less than 4000 lines
of code, but this simplicity also means that there are a number of limitations:
\begin{itemize}
  \item Only second-order accurate 5/7-point discretizations of elliptic
  operators are supported for now.
  \item Polar and spherical grids are not supported. They are not compatible
  with the point-wise smoothers used here, see section
  \ref{sec:smoother}.
  \item Geometric multigrid is here used as a solver. With libraries such as
  PETSc and Hypre multigrid can also be used as a preconditioner.
  \item Strong scaling is here demonstrated up to about $2 \times 10^3$
  processors. For significantly larger runs, a more sophisticated parallel
  implementation could be required, see section \ref{sec:parallelization}.
  Multigrid methods that are potentially more suitable for such large problems
  can be found in e.g.~\cite{Gholami_2016,Adams_2014,falgout:hypre}.
\end{itemize}

\emph{Contents of the paper:} The design and implementation of the library are
described in section \ref{sec:geom-mult-libr}. Afterwards, several convergence
and scaling tests are presented in section \ref{sec:testing-library}. Finally,
we use the library section for divergence cleaning in MHD simulations with
\texttt{MPI-AMRVAC} in section \ref{sec:divergence-cleaning}.

\section{Geometric multigrid library}
\label{sec:geom-mult-libr}

\subsection{Introduction to multigrid}
\label{sec:mg-intro}

Below, we provide only a brief introduction to multigrid methods. For a more
detailed introduction to multigrid, there exist a number of excellent textbooks
and review papers, see for
example~\cite{Hackbusch_1985,trottenberg_2000_multigrid,briggs_2000,brandt_2011}.

Relaxation methods such as the Gauss-Seidel method and successive
over-relaxation (SOR) can be used to solve elliptic PDEs. However, the
convergence rate of such methods decreases for larger problem sizes, which can
be analyzed by decomposing the error into different wavelengths. Typically, only
short wavelength errors are effectively damped. One reason for this is that the
solution is locally updated, so that it can take a large number of iterations
for information to propagate throughout the domain. Because relaxation methods
locally smooth the error, they are also referred to as \emph{smoothers}. The
multigrid library presented here includes a couple relaxation methods /
smoothers, which are described in section \ref{sec:smoother}.

The main idea behind geometric multigrid methods is to accelerate the
convergence of a relaxation method by applying it on a hierarchy of grids. With
the relaxation method, short wavelength errors can effectively be damped on any
grid in the hierarchy. However, a short wavelength on a coarse grid corresponds
to a long wavelength on a fine grid. By combining information from all grid
levels, it is possible to efficiently damp all of the error wavelengths. The
transfer of information between grid levels is done by prolongation
(interpolation) and restriction, which are described in section
\ref{sec:restr-prol}. The order in which relaxation, prolongation and
restriction are performed is determined by the multigrid cycle type. We support
two popular options, namely V-cycles and full multigrid (FMG) cycles, see section
\ref{sec:multigrid-cycles}.

\subsection{Included operators and smoothers}
\label{sec:smoother}

When an elliptic PDE is discretized on a mesh with grid spacing $h$, it can be
written in the following form
\begin{equation}
  \label{eq:discretization}
  L^h \phi^h = f^h,
\end{equation}
where $L$ is an elliptic operator, $\phi$ is the solution to be obtained, $f$ is
the right-hand side, and the superscript $h$ indicates that these quantities are
discretized. The multigrid library contains several predefined elliptic
equations, namely:
\begin{itemize}
  \item $\nabla \cdot (\nabla \phi) = f$: Poisson's equation
  \item $\nabla \cdot (\varepsilon \nabla \phi) = f$: Poisson's equation
  with a variable coefficient $\varepsilon$
  \item $\nabla \cdot (\nabla \phi) - \lambda \phi = f$: Helmholtz
  equation with $\lambda \geq 0$
  \item $\nabla \cdot (\varepsilon \nabla \phi) - \lambda \phi = f$: Helmholtz
  equation with a variable coefficient $\varepsilon$
\end{itemize}
These equations are discretized with a standard 5/7-point second-order accurate
discretization, in which the solution and the right-hand side are defined at
cell centers. The library supports grids with structured adaptive mesh
refinement, see section~\ref{sec:grid-structure}. On such grids, the
discretization of Poisson's equation in 2D at a cell $(i,j)$ is for example
given by
\begin{equation}
\label{eq:poisson-2d}
h_x^{-2} (\phi_{i-1,j} - 2 \phi_{i,j} + \phi_{i+1,j})
+ h_y^{-2}(\phi_{i,j-1} - 2 \phi_{i,j} + \phi_{i,j+1}) = f_{i,j},
\end{equation}
where $h_x$ and $h_y$ denote the grid spacing in the $x$ and $y$
directions, respectively. The generalization to 3D is straightforward (an extra
term for the $z$-direction appears). With a variable coefficient, we use the
following discretization for Poisson's equation in 2D
\begin{align}
    \label{eq:poisson-2d-vcoeff}
  &h_x^{-2} \left[
  \bar{\varepsilon}_{i-1/2,j} (\phi_{i-1,j} - \phi_{i,j}) +
  \bar{\varepsilon}_{i+1/2,j} (\phi_{i+1,j} - \phi_{i,j})\right]\\
  &+ h_y^{-2} \left[
  \bar{\varepsilon}_{i,j-1/2} (\phi_{i,j-1} - \phi_{i,j}) +
  \bar{\varepsilon}_{i,j+1/2} (\phi_{i,j+1} - \phi_{i,j})\right] = f_{i,j},
\end{align}
where $\bar{\varepsilon}$ denotes the harmonic mean of the coefficients in
neighboring cells. For example, the coefficient between cells $(i-1, j)$ and
$(i, j)$ is defined as
\begin{equation*}
  \bar{\varepsilon}_{i-1/2,j} = \frac{2 \, \varepsilon_{i,j} \,
    \varepsilon_{i-1,j}}{\varepsilon_{i,j} + \varepsilon_{i-1,j}}.
\end{equation*}
The variation in the coefficients has to be smooth enough for standard multigrid
methods to work, since we adopt no special treatment for discontinuous
coefficients (see for example~\cite{brandt_2011}).

Besides the equations listed above, users can also define their own elliptic
operators. Currently, the library supports discrete operators with 5-point
stencils in 2D and 7-point stencils in 3D. The advantage of employing such
sparse stencils (without diagonal elements) is that the amount of communication
between processors is significantly reduced. However, the library could
relatively easily be extended to support 9/27-point stencils in 2D/3D that also
use diagonal elements.

When performing multigrid, a smoother (relaxation method) has to be employed to
smooth the error in the solution. We include point-wise smoothers of the
Gauss-Seidel type, which solve the discretized equations for $\phi_{i,j}$ while
keeping the values at neighbors fixed. For example, for equation
\eqref{eq:poisson-2d} the local solution $\phi_{i,j}^*$ is given by
\begin{equation*}
  \phi_{i,j}^* = \frac{1}{h_x^{-2} + h_y^{-2}}
  \left[
    h_x^{-2} (\phi_{i-1,j} + \phi_{i+1,j})
    + h_y^{-2}(\phi_{i,j-1} + \phi_{i,j+1}) - f_{i,j} \right].
\end{equation*}
The order in which a smoother replaces the old values $\phi_{i,j}$ by
$\phi_{i,j}^*$ affects the smoothing behavior. Two orderings are provided:
\begin{itemize}
  \item Standard Gauss-Seidel, which linearly loops over all the $(i, j)$
  indices (in the order they are stored in the computer's memory).
  \item Gauss-Seidel red--black, which first updates all points for which $i + j$
  is even, and then all points for which $i + j$ is odd.
\end{itemize}

A downside of point-wise smoothers is that they require the `coupling' between unknowns
to be of similar strength in all directions, otherwise the convergence rate is
reduced (see e.g.~\cite{briggs_2000}). For a Laplace equation
$\nabla^2 \psi = 0$ on a Cartesian grid, this means that $h_x$, $h_y$ and $h_z$
have to be similar, e.g.~within a factor two. This also restricts the geometries
in which a point-wise smoother can be applied. For example, in 3D cylindrical
coordinates the Laplace equation becomes
$$
\frac{1}{r}\partial_r (r \partial_r \psi) + \frac{1}{r^2} \partial_\phi^2 \psi + \partial_z^2 \psi = 0.
$$
The $1/r^2$ factor in front of the $\partial_\phi^2$ term violates the
similar-coupling requirement, which is why 3D cylindrical coordinates are not
supported in the library. However, a discretization for a constant-coefficient
Poisson equation in a 2D axisymmetric geometry is provided.

So-called line smoothers or plane smoothers solve for multiple unknowns along a
line or a plane simultaneously. They are typically more robust than point-wise
smoothers, and they can be used to perform multigrid in polar/spherical
coordinate systems~\cite{Barros_1988,Barros_1991}. However, line or plane
smoothers are incompatible with grid refinement if standard multigrid cycles are
used, because they would have to solve for unknowns at different refinement
levels.

\subsection{Prolongation and restriction}
\label{sec:restr-prol}

\begin{figure}
  \centering
  \includegraphics[width=5.0cm]{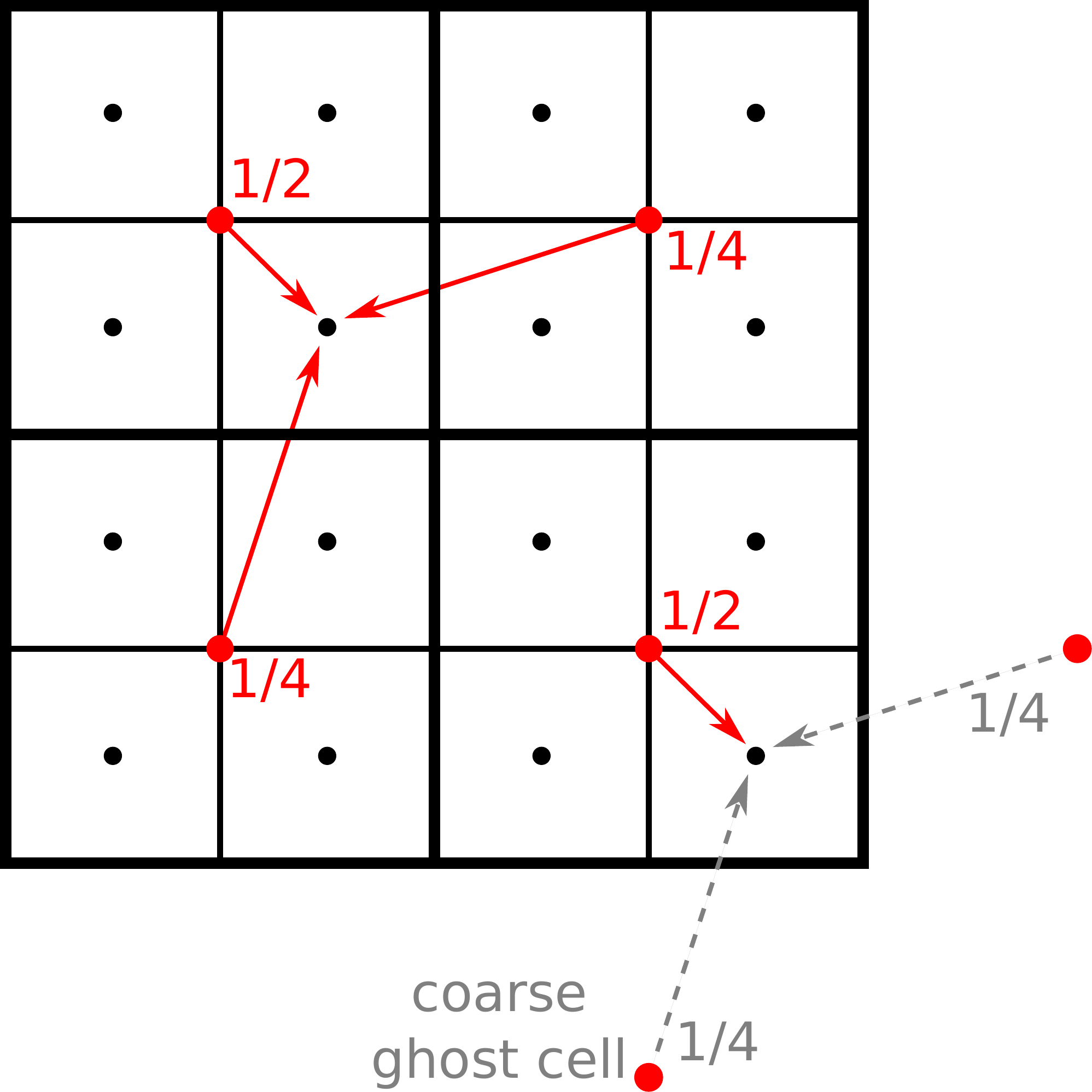} %
  \caption{Illustration of the prolongation procedure in 2D. Cell-centered
    values on the coarse and fine grids are indicated by red and black circles,
    respectively. Arrows and weights indicate which coarse grid values are used
    for the interpolation. Near block boundaries, the interpolation makes use of
    coarse grid ghost cells, but note that no diagonal ghost cells are used.
    Formulas for the interpolation scheme are given in equations
    \eqref{eq:interp_2d} and \eqref{eq:interp_3d}.}
  \label{fig:interp-2d}
\end{figure}

Besides a smoother, multigrid also requires prolongation and restriction
methods, which transfer data from coarse to fine grids and vice versa. For
prolongation we use linear interpolation based on the nearest neighbors, as is
also included in e.g.~the Boxlib~\cite{Zhang_2016} and
Afivo~\cite{Teunissen_2018_afivo} frameworks. The procedure is illustrated in
figure \ref{fig:interp-2d} for a 2D case, and can be described by the following
equations:
\begin{align}
  \label{eq:interp_2d}
  f_{x+h/4, y+h/4} &= \frac{1}{4} \left(2 f_{x,y} + f_{x+h,y} + f_{x,y+h}\right) + O(h^2),\\
  f_{x-h/4, y+h/4} &= \frac{1}{4} \left(2 f_{x,y} + f_{x-h,y} + f_{x,y+h}\right) + O(h^2),\nonumber
\end{align}
with the schemes for other points following from symmetry. In 3D, the
interpolation stencil becomes
\begin{align}
  \label{eq:interp_3d}
  f_{x+h/4, y+h/4,z+h/4} &= \frac{1}{4} \left(f_{x,y,z} + f_{x+h,y,z} + f_{x,y+h,z} +
                           f_{x,y,z+h}\right) + O(h^2),\\
  f_{x-h/4, y+h/4,z+h/4} &= \frac{1}{4} \left(f_{x,y,z} + f_{x-h,y,z} + f_{x,y+h,z} +
                           f_{x,y,z+h}\right) + O(h^2).\nonumber
\end{align}
An advantage of these schemes is that they do not require diagonal ghost
cells, which saves significant communication costs. A drawback is that
interpolation errors can be larger than with standard bilinear or trilinear
interpolation, somewhat reducing the multigrid convergence rate.

For restriction, the value of four (2D) or eight (3D) fine grid values is
averaged to obtain a coarse grid value. Besides these built-in methods, users
can also define custom prolongation and restriction operators.

\subsection{Multigrid cycles}
\label{sec:multigrid-cycles}

\begin{figure}
  \centering
  \includegraphics[width=8.0cm]{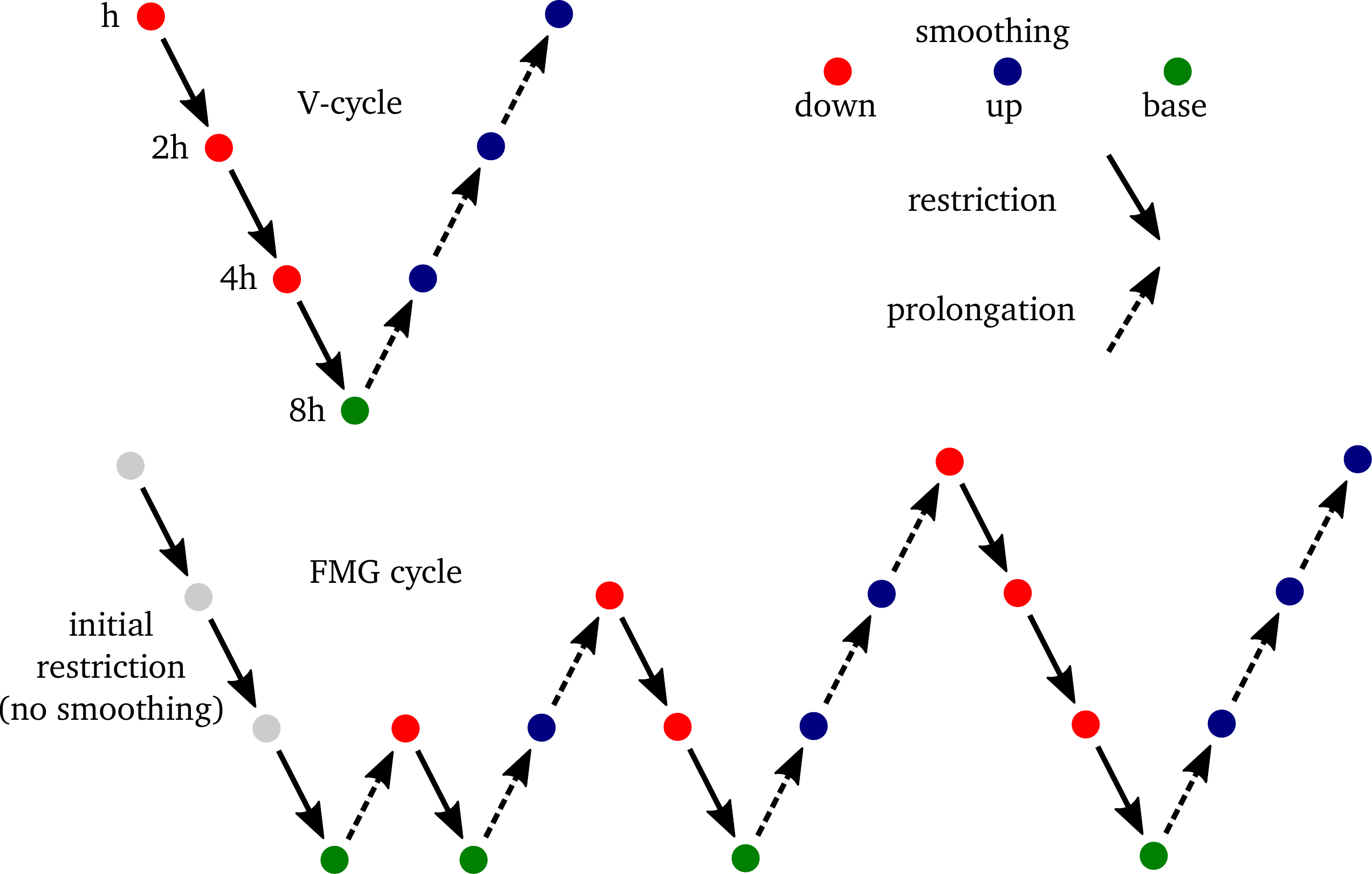} %
  \caption{Schematic illustration of the V-cycle and FMG cycle for a grid with
    four levels. The FMG cycle contains several V-cycles at increasingly finer
    grids. At the start of an FMG cycle, the solution and the right-hand side
    are restricted to the coarsest grid.}
  \label{fig:mg-cycles}
\end{figure}

Two standard multigrid cycles are included in the
library~\cite{brandt_2011,briggs_2000,trottenberg_2000_multigrid}: the V-cycle
and the full multigrid (FMG) cycle, which are illustrated in figure
\ref{fig:mg-cycles}. As the name suggests, a V-cycle goes from the finest grid
to the coarsest grid and then back to the finest grid. To explain the procedure,
we introduce some terminology: let $v$ denote the approximate solution, $f$ the
right-hand side, $L$ the elliptic operator, $r = f - L v$ the residual, $P$ a
prolongation operator, and $R$ a restriction operator. Furthermore, superscripts
$h$ and $H$ refer to the current and the underlying coarse grid level.

During the downward part of the V-cycle, $N_\mathrm{down}$ (default: two)
smoothing steps are performed at a grid level. Afterwards, the residual is
computed as
$$r^h = f^h - L^h(v^h).$$
The current approximation $v^h$ is then restricted to the coarse grid as
$v^H = R(v^h)$, after which a copy $v^H_{\mathrm{old}} = v^H$ is stored. This
copy is later used to update the fine-grid solution. The coarse-grid right-hand
side is then updated as
\begin{equation}
  f^H = R(r^h) + L^H(v^H),
  \label{eq:coarse-rhs}
\end{equation}
after which the procedure repeats itself, but now starting from the underlying
coarse grid.

On the coarsest grid, up to $N_\mathrm{max}$ (default: 1000) smoothing steps are
performed until the residual is either below a user-defined absolute threshold
(default: $10^{-8}$), or until it is reduced by a user-defined factor (default:
$10^{-8}$). When the coarsest grid contains a large number of unknowns, it can
be beneficial to use a direct solver to solve the coarse grid equations, but we
have not yet implemented this.

In the prolongation steps, the solution is updated with a correction from the
coarse grid as
\begin{equation}
  v^h = v^h + P(v^H - v^H_{\mathrm{old}}),
  \label{eq:coarse-corr}
\end{equation}
and afterwards $N_\mathrm{up}$ (default: two) smoothing steps are performed.

The FMG cycle consists of a number of V-cycles, as illustrated in figure
\ref{fig:mg-cycles}. Compared to V-cycles, FMG cycles perform additional
smoothing at coarse grid levels. This makes them a bit more expensive, but the
advantage of FMG cycles is that they can achieve convergence up to the
discretization error in one or two iterations.

If no initial guess for the solution is given, an initial guess of zero is used.
We use the restriction and prolongation operators described in section
\ref{sec:restr-prol} for both V-cycles and FMG cycles. Whenever necessary, for
example after restriction/prolongation or after a smoothing step, ghost cells
are updated, see section \ref{sec:ghost-cells-boundary}.

\subsection{Grid structure}
\label{sec:grid-structure}

\begin{figure}
  \centering
  \includegraphics[width=8cm]{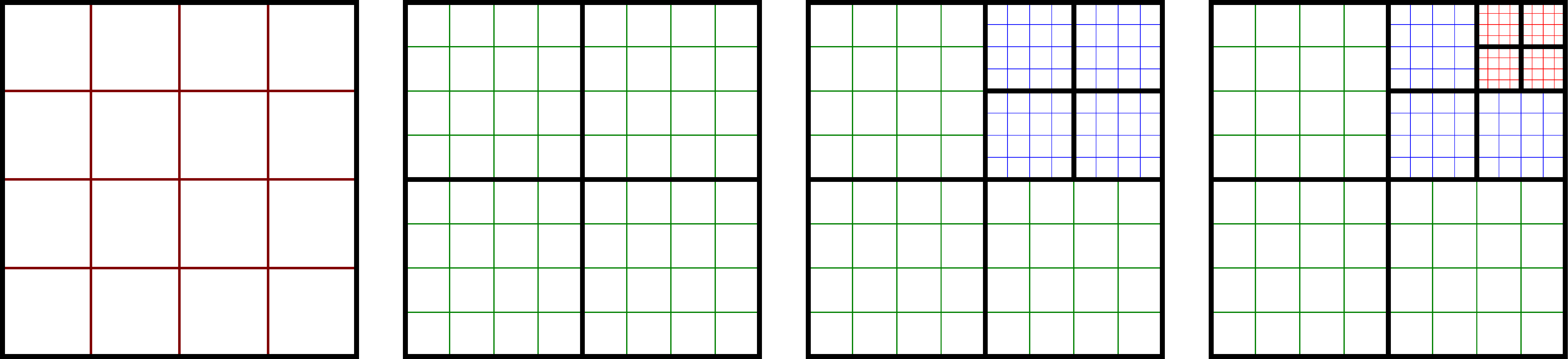} %
  \caption{Example of a quadtree grid. Each black square represents a grid block
    containing $4 \times 4$ cells. From left to right, the grid is refined in
    the upper right corner.}
  \label{fig:quadtree}
\end{figure}

The library supports quadtree/octree grids with Cartesian 2D/3D or cylindrical
2D $(r,z)$ geometries, see e.g.~\cite{keppens_2012}. Quadtree grids consist of
blocks of $N_x \times N_y$ cells, which can be refined by covering them with
four refined blocks (their `children'), which each also contain $N_x \times N_y$
cells but have half the grid spacing. An example of a quadtree grid is shown in
figure \ref{fig:quadtree}. Octrees are the 3D equivalent of quadtrees, so that
the refinement of a block creates eight `children'. The multigrid library
requires that the difference in refinement between adjacent blocks is at most
one level; such quadtree/octree grids are called \emph{2:1 balanced}.

When the multigrid library is coupled to an application code, it constructs a
copy of the full AMR hierarchy of the application code together with additional
coarse grid levels. Applications codes therefore only need to contain fine grid
data, and not the underlying coarse grid data. The library also contains its own
routines for parallel communication and the filling of ghost cells, as described
in sections \ref{sec:ghost-cells-boundary} and \ref{sec:parallelization}.

The grid construction is performed in several steps. First, a user indicates the
quadtree/octree block size and the size of the unrefined computational domain in
the application code (in number of cells). The library will then internally
construct additional coarse grid levels, as illustrated below. Afterwards, the
refinement levels present in the application code are copied to the multigrid
library.

\subsubsection{Construction of additional coarse grids}

\begin{figure}
  \centering
  \includegraphics[width=8cm]{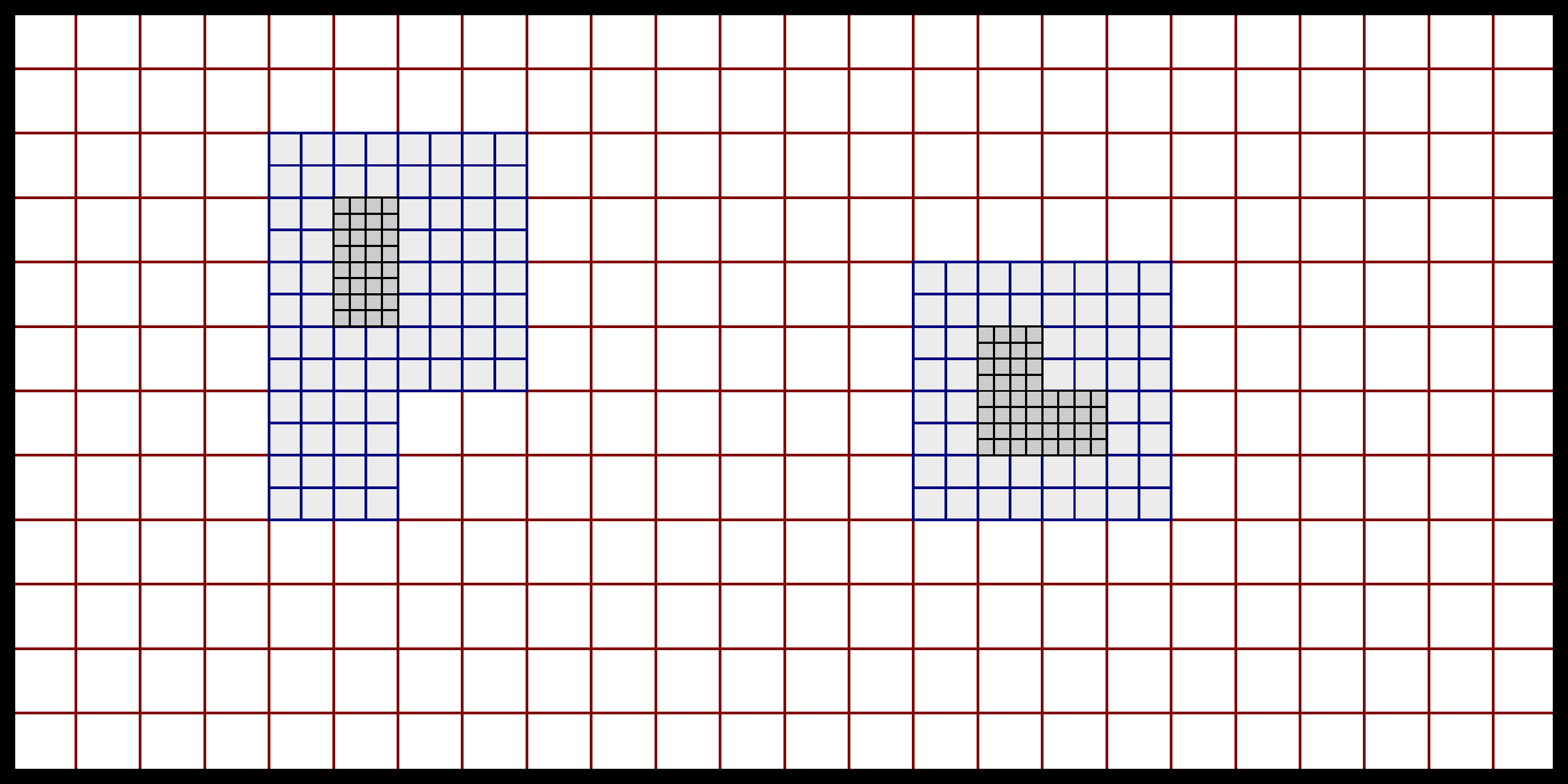}
  \caption{Example of a 2D grid with two levels of refinement, to help explain
    the construction of grids in the multigrid library. The base grid contains
    $192 \times 96$ cells, which corresponds to $24 \times 12$ blocks (shown in
    red) of $8 \times 8$ cells. Table \ref{tab:coarse-grid-example} lists the
    additional coarse grids that the multigrid library would construct.}
  \label{fig:2d-grid-construction}
\end{figure}

Suppose that a 2D application uses blocks of $8^2$ cells and that its level one
grid contains $192 \times 96$ cells, see figure \ref{fig:2d-grid-construction}.
The library will then first construct the additional coarse grids given in table
\ref{tab:coarse-grid-example}. The block size is kept at $8 \times 8$ down to
level $-1$. For levels $-2$ to $-4$, the block size is reduced all the way down
to $1\times 1$ blocks. These extra grids are constructed to obtain a coarsest
grid with a small number of unknowns. On such a grid, a solution can directly be
obtained with a modest number of iterations of the smoother. A coarsest grid with
few unknowns can be obtained when the level one grid size is a small number
(e.g., 1, 3 or 5) times a power of two.

\begin{table}[!h]
  \centering
  \small
  \begin{tabular}{cccccc}
    grid level & mg & application & grid size & block size & $N_\mathrm{blocks}$\\
    \hline
    3 & \checkmark & \checkmark & irregular & $8\times 8$ & 80 \\
    2 & \checkmark & \checkmark & irregular & $8\times 8$ & 144 \\
    \hline
    1 & \checkmark & \checkmark & $192 \times 96$ & $8\times 8$ & $24 \times 12$ \\
    0 & \checkmark &  & $96 \times 48$ & $8\times 8$ & $12 \times 6$ \\
    -1 & \checkmark &  & $48 \times 24$ & $8\times 8$ & $6 \times 3$ \\
    \hline
    -2 & \checkmark &  & $24 \times 12$ & $4\times 4$ & $6 \times 3$\\
    -3 & \checkmark &  & $12 \times 6$ & $2\times 2$ & $6 \times 3$\\
    -4 & \checkmark &  & $6 \times 3$ & $1\times 1$ & $6 \times 3$
  \end{tabular}
  \caption{Example of the additional coarse grids that would be constructed for
    an unrefined domain of $192 \times 96$ cells with blocks of size $8^2$, see
    figure \ref{fig:2d-grid-construction}. The check marks indicate whether the
    grid level is present in the multigrid library and in the calling
    application. $N_\mathrm{blocks}$ denotes the number of blocks per level in
    the multigrid hierarchy, so including blocks covered by refinement.}
  \label{tab:coarse-grid-example}
\end{table}

\subsubsection{Copying the application's grid refinement}

After the unrefined (level one) grid has been constructed in the multigrid
library, it can be linked to the application's unrefined grid. In the
application code, each grid block has to store an integer indicating the index
of that grid block in the multigrid library. Similarly, grid blocks in the
multigrid library store pointers to the application's code grid blocks.

After the unrefined grids have been linked, the AMR structure can be copied from
the application code by looping over its grid blocks, starting at level one.
Refined blocks can be added to the multigrid library by calling built-in
refinement procedures, after which these refined blocks again have to be linked
between the two codes. An example of the coupling procedure can be found in the
coupling module provided for \texttt{MPI-AMRVAC}.

\subsubsection{Adapting the grid structure}

When the mesh in the calling application changes, the mesh in the multigrid
library can be adapted in the same way, or it can be constructed again from
scratch. To adapt an existing mesh the calling application should inform the
library about all blocks that were added, removed or transferred between
processors (for load balancing, see section \ref{sec:parallelization}). The
computational cost of constructing a new mesh is relatively modest: for the
uniform-grid scaling tests in section \ref{sec:scaling-tests}, it took about
$0.3 \, \textrm{s}$ to construct a $1024^3$ grid consisting of octree blocks
with $16^3$ cells.

\subsection{Ghost cells and boundary conditions}
\label{sec:ghost-cells-boundary}

In the multigrid library all grid blocks have a layer of ghost cells around
them, which can contain data from neighboring blocks (potentially at a different
refinement level) or special values for boundary conditions. The usage of ghost
cells simplifies the implementation of numerical methods, since they do not need
to take block boundaries into account. For simplicity and efficiency, the
library currently uses only a single layer of ghost cells, without diagonal
and/or edge (in 3D) cells. Since the multigrid library uses its own ghost cell
routines, these restrictions do not apply to application codes, which can use
any number of ghost cells.

The downside of ghost cells is that additional memory is required, as
illustrated in table \ref{tab:ghost-cell-cost}. Some AMR codes, such as
\texttt{Paramesh}~\cite{macneice_2000}, therefore provide the possibility to
compute ghost values when they are required instead of permanently storing them.
However, this adds some complexity in the implementation of algorithms, for
example because ghost cells cannot be reused in two separate steps.

\begin{table}
  \centering
  \begin{tabular}[center]{c|ccc}
    block size & 1 ghost cell & 2 ghost cells & 3 ghost cells\\
    \hline
    $8^2$ & 1.56 & 2.25 & 3.06\\
    $16^2$ & 1.27 & 1.56 & 1.89\\
    $32^2$ & 1.13 & 1.27 & 1.41\\
    \hline
    $8^3$ & 1.95 & 3.38 & 5.36\\
    $16^3$ & 1.42 & 1.95 & 2.60\\
    $32^3$ & 1.20 & 1.42 & 1.67
  \end{tabular}
  \caption{Memory cost of using grid blocks of given size with ghost cells,
    relative to the cost without ghost cells. The values are computed as
    $(N + 2 N_\mathrm{gc})^D / N^D$, where $N$ is the block size,
    $N_\mathrm{gc}$ the number of ghost cells, and $D$ the problem dimension.}
  \label{tab:ghost-cell-cost}
\end{table}

Ghost cells can be filled in three different ways. If there is a neighboring
block (at the same refinement level), ghost cells are simply copied from the
corresponding region. This is also performed at periodic boundaries.

\paragraph{Ghost cells near physical boundaries}

If there is a physical boundary, ghost cells are set so that the boundary
condition is satisfied at boundary cell faces. If the interior cell-centered
value is $\phi_i$ and the ghost value is $\phi_g$, then a Dirichlet boundary
condition $\phi = a$ at the cell face is set as $\phi_g = 2 a - \phi_i$. A
Neumann boundary condition $\partial_x \phi = b$ is set as
$\phi_g = \phi_i \pm h_x \, b$, where $h_x$ is the grid spacing and the sign
depends on the direction the boundary is facing. For free space boundary
conditions ($\phi \to 0$ for $r \to \infty$), we make use of a FFT-based solver
to set boundary conditions, see section \ref{sec:free-space-boundary}.

\begin{figure*}
  \centering
  \includegraphics[width=14.0cm]{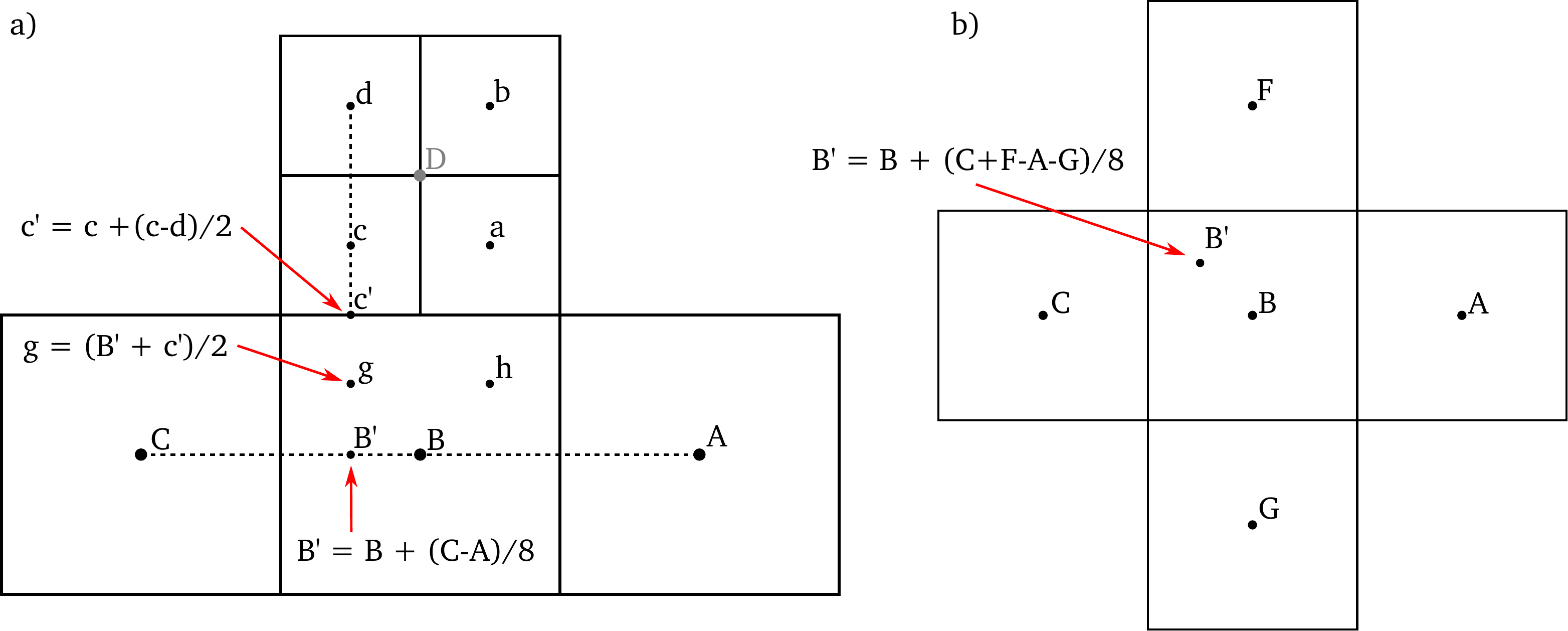} %
  \caption{a) Illustration of the ghost cell scheme near a refinement boundary
    in 2D. Fine-grid values are indicated by a to d. The ghost cell next to c is
    located at g, and the nearest coarse-grid value is indicated by B. The
    quantities at B' and c' are shown to help explain how a value for g is
    obtained. Note that the coarse values A or C can be inside ghost cells of
    the block containing B; values at such ghost cells are always available in
    our implementation. b) In 3D, the procedure is almost identical. The only
    difference is how the equivalent of B' is determined, which is illustrated
    here.}
  \label{fig:refinement-boundary}
\end{figure*}

\paragraph{Ghost cells near refinement boundaries}

Near refinement boundaries, we employ the scheme illustrated in figure
\ref{fig:refinement-boundary}a to fill ghost cells. The value B' is obtained
by using the central-difference slope in the coarse grid cell. The value at the
cell face c' is obtained by local extrapolation using two points, and finally
the ghost cell value g is the average of B' and c'. For the other ghost
value h the procedure is geometrically identical. The approach extends
naturally to 3D, in which two central-difference slopes are used in the coarse
cells to obtain the equivalent of B', as illustrated in figure \ref{fig:refinement-boundary}b. An important
property of this ghost cell scheme, which is similar to the scheme presented
in~\cite{Teunissen_2018_afivo}, is that it gives the same coarse and (average)
fine gradient across the refinement boundary. In other words, we have that
\begin{equation}
  D - B = (c-g)/2 + (a-h)/2,
\end{equation}
where $D = (a + b + c + d)/4$ is the restriction of the fine cells. For a
constant-coefficient Poisson equation such as \eqref{eq:poisson}, the divergence
theorem
\begin{equation}
  \int f \, dV = \oiint \varepsilon \nabla \phi \cdot \vec{dS}.
\end{equation}
then shows that the integrated right-hand side is equal on the refined patch and
on the underlying coarse grid approximation.

Near refinement boundaries, ghost cell values depend on the interior values of
the refined block, as illustrated in figure \ref{fig:refinement-boundary}. Since
ghost cells are updated after (and not during) a smoothing step, a somewhat
slower damping of errors is to be expected near refinement boundaries.

\subsection{Parallelization}
\label{sec:parallelization}

We have made a number of choices to keep the parallel implementation of the
multigrid library relatively simple.

First, the full mesh geometry is stored on every processor. Of course, each
processor only allocates storage for the mesh data that it `owns'. A more
sophisticated approach would be to only store information about the local mesh
neighborhood for each processor. That would save memory, but also significantly
complicate e.g.~mesh construction, mesh adaptation and load balancing.

Second, the multigrid library copies the mesh structure from the calling AMR
application, and it is assumed that 2:1 balance is already satisfied. 

Third, the load balancing is also copied from the calling AMR application, which
means that all leaves (i.e., blocks with no further refinement) are on the same
processors in the library as in the calling application. Copying data between
the calling application and the library on the leaf blocks therefore involves no
communication. This also means that the library only needs to perform load
balancing for parent blocks. Our implementation assigns each parent block to the
processor that contains most of its children, which is applied recursively by
going from finer to coarser grids. In case of ties, the processor which has the
fewest blocks at a refinement level is selected. This type of load balancing
minimizes the communication between children and parent blocks.

On the coarsest grids in a multigrid hierarchy, there are too few unknowns to
keep all processors busy. Furthermore, the cost of communication on such grids
is often higher than computational costs. We therefore store the coarsest grids
on a single processor, more precisely those for which the number of cells is not
divisible by the block size. For the example of table
\ref{tab:coarse-grid-example}, these are the grids with block size $4\times 4$
and smaller.

\paragraph{Parallel communication}

As illustrated in figure \ref{fig:mg-cycles}, performing a multigrid cycle
involves quite a lot of communication between processors. After performing a
smoothing step, ghost cells have to be updated. Prolongation and restriction also
require data to be transferred between grid levels, as well as an update of the
ghost cells. For this reason, the multigrid library comes with efficient
routines for filling ghost cells and communicating data.

The following data is communicated for the ghost cell, prolongation and
restriction routines:
\begin{itemize}
  \item For ghost cell exchanges at the same refinement level, the corresponding
  interior cell region is sent from both sides.
  \item For ghost cells near a refinement boundary, the coarse-side processor
  interpolates values `in front of' the cells of its fine grid neighbor, see
  figure \ref{fig:refinement-boundary}. These values are then sent from coarse
  to fine; there is no communication from fine to coarse.
  \item For prolongation, the coarse grid data is first interpolated and then
  sent to its children\footnote{It is more efficient to send coarse data and
    to perform interpolation afterwards, but this approach is less flexible; for
    a variable coefficient problem, it can for example be beneficial to change
    the interpolation scheme depending on the local coefficients.}.
  \item For restriction, the fine grid data is first restricted and then sent to
  the underlying coarse grid.
\end{itemize}
For each of the above cases, the size of the data transferred depends only on the
block size and the problem dimension. We avoid communication on the coarsest
grids, for which the block size is reduced, since these grids are stored on a
single processor.

The actual data transfer is performed using buffers, so that only a single send
and/or receive is performed between communicating processors. The size of these
buffers is computed after constructing the AMR grid; then it is known how much
data is sent and received between processors in the various operations.
Furthermore, the data in the send buffers is sorted so that it is in `natural'
order for the receiving processor. This sorting is possible because each grid
block is identified by a global index, which determines the order in which
processors loop over the blocks.

After the sorted data has been received, operations such as the filling of ghost
cells, prolongation or restriction can be performed. Whenever data from another
processor is required, it is unpacked from the buffer corresponding to that
processor. If data from the same processor is required, it is locally prepared as
described above. The advantage of this buffered approach for exchanging data is
that it limits the number of MPI calls, which could otherwise lead to
significant overhead. In section \ref{sec:scaling-tests}, we demonstrate the
parallel scaling of our approach.

\subsection{Free space boundary conditions in 3D}
\label{sec:free-space-boundary}

Poisson's equation sometimes has to be solved with \emph{free space} boundary
conditions, i.e., $\phi \to 0$ at infinity, for example when computing the
gravitational potential of an isolated system. Because of the $1/r$ decay
of the free-space Green's function, enlarging the computational domain (and
applying a Dirichlet zero boundary condition) gives a poor approximation of
the free-space solution.

A number of techniques exist to directly compute free-space solutions, see e.g.
\cite{genovese_2006,Hejlesen_2013}. The most efficient techniques rely on the
fast Fourier transform (FFT), so they can only be applied to uniform grids. To
incorporate free boundary conditions into our AMR-capable multigrid solver, we
therefore employ the following strategy. First, a free-space solution is
computed on a uniform grid, which can have a significantly lower resolution than
the full AMR grid. Then standard multigrid is performed, using Dirichlet
boundary conditions interpolated from the uniform grid solution.

We use the 3D uniform-grid solver described in
\cite{genovese_2006,Genovese_2007}, which employs interpolating scaling
functions and FFTs to obtain high-order solutions of free-space problems. The
solver is written in Fortran, licensed under a GPL license, and it uses
MPI-parallelism, which simplified its integration with our multigrid library.

In our implementation the uniform grid always corresponds to one of the fully
refined grid levels (so excluding partially refined levels). Users can control
the cost of the uniform grid solver with a parameter $c$, which should lie
between zero and one. The uniform grid then corresponds to AMR level $l$ for
which $N(l) \leq c \, N_\mathrm{total}$, where $N(l)$ denotes the number of
unknowns at AMR level $l$, and $N_\mathrm{total}$ denotes the total number of
unknowns.

After constructing the uniform grid, including a layer of ghost cells, the
right-hand side of the problem is restricted to it. The direct solver then
computes the free-space solution in parallel, using eight-order accurate
interpolating scaling functions. We extract the boundary planes, and linearly
interpolate them to obtain Dirichlet boundary conditions for the multigrid
solver at all grid levels. Afterwards, one or more FMG or V-cycles can be
performed to obtain a solution on the full AMR grid, using the uniform grid
solution as an initial guess.


The coupled approach described above has two advantages: it can handle AMR
grids, it can be more efficient and scale better than a direct solver, and it
requires no modification of the multigrid routines. Potential drawbacks are that
the multigrid solution is only second order accurate, and that the accuracy near
boundaries is reduced when a too coarse grid is used for the direct solver. How
fine the uniform grid needs to be compared to the full AMR grid depends on the
application, e.g., on the distance between sources and the domain boundary, and
on the required accuracy near the boundary.




\section{Testing the library}
\label{sec:testing-library}

\subsection{Convergence test}
\label{sec:convergence-test}

To study the convergence behavior of the multigrid solver, we solve the
following 3D test problem on a unit cube centered at the origin:
\begin{align}
  \label{eq:poisson-conv}
  \nabla^2 \phi &= f,\nonumber\\
  f &= \nabla^2 \phi_\mathrm{sol},\\
  \phi_\mathrm{sol} &= \cos(\pi \vec{n} \cdot \vec{x}) + 10 \exp(-100 |\vec{x}|^2),\nonumber
\end{align}
with $\vec{n} = (1, 2, 3)$. The right-hand side $f$ is computed analytically,
and Dirichlet boundary conditions are imposed using the solution values at the
boundary. We consider three types of numerical grid. The base case has uniform
refinement using $64^3$ cells. To test the effect of refinement boundaries on
the convergence behavior, we add two extra levels of refinement covering a
volume of $0.5^3$ and $0.25^3$, respectively (so that each level again contains
$64^3$ cells). These refinements are either placed at the center of the domain,
or around $(-0.25, -0.25, -0.25)$. In the latter case, the refinement is in the
`wrong' place, as it leads to a refinement corner at the center of the domain,
where the solution has a sharp peak.

\begin{figure*}
  \centering
  \includegraphics[width=6.5cm]{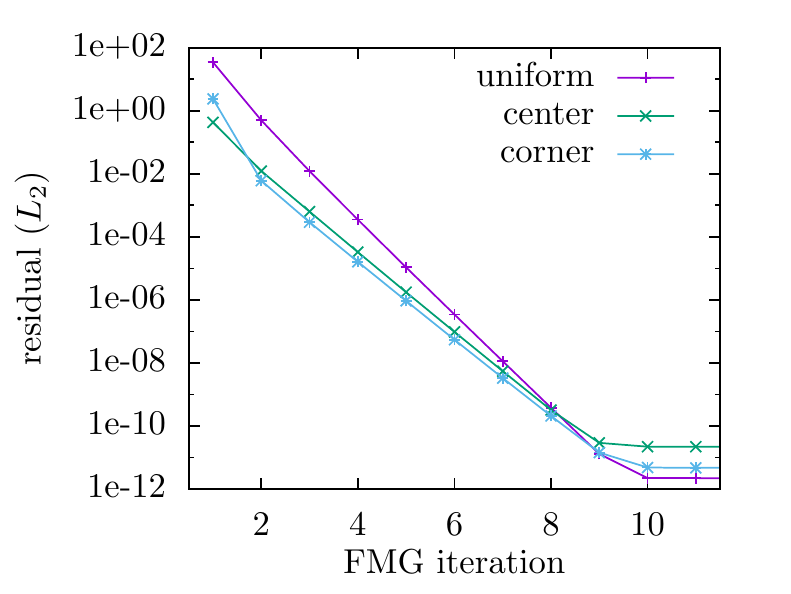}%
  \includegraphics[width=6.5cm]{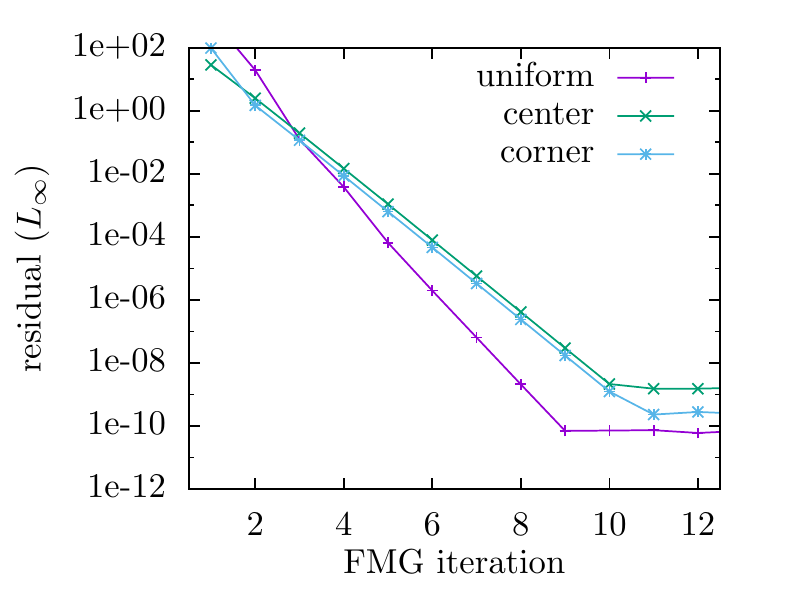}
  \includegraphics[width=6.5cm]{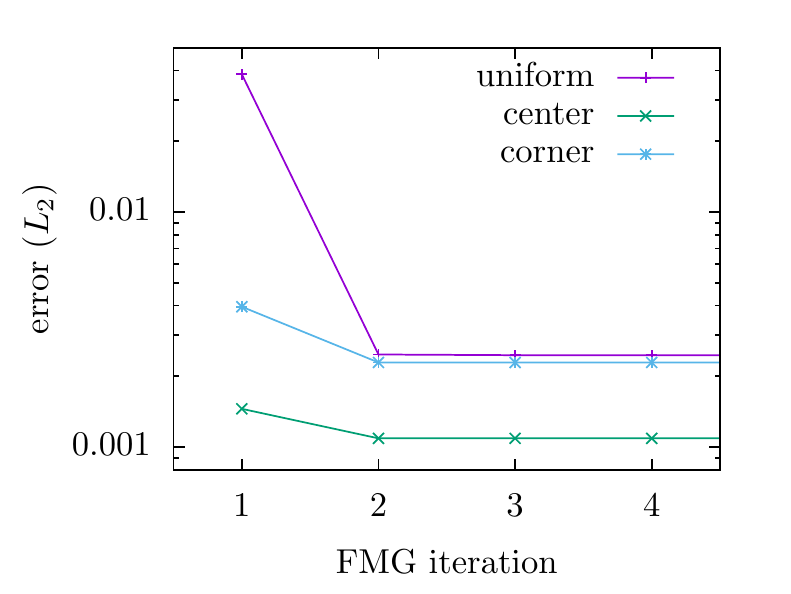}%
  \includegraphics[width=6.5cm]{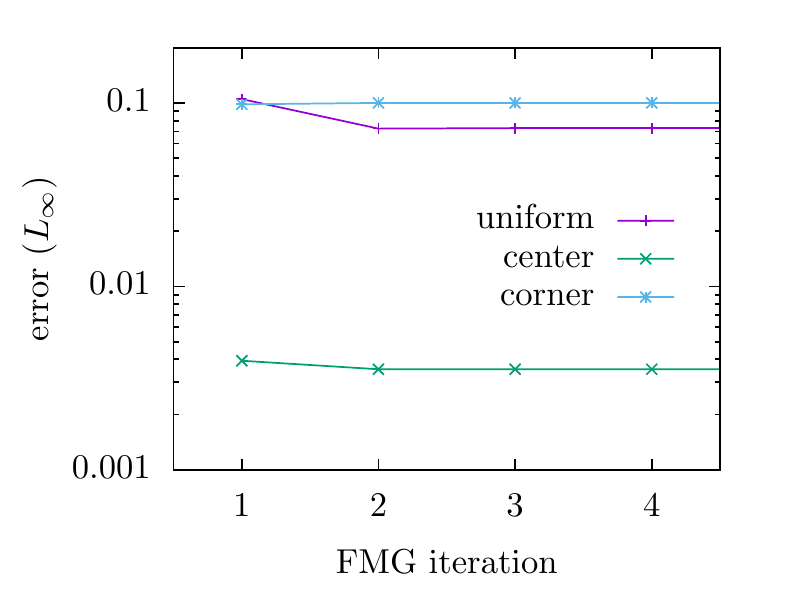}
  \caption{Convergence test results for the test problem of equation
    \eqref{eq:poisson-conv} with a sharp Gaussian peak. Top row: the residual in
    $L_2$ norm (left) and $L_\infty$ norm (right) versus FMG iteration. After
    about 10 iterations the residual is reduced up to machine precision, see the
    text for details. Bottom row: the $L_2$ norm and $L_\infty$ norm of the
    solution error. After two FMG iterations, the error has converged to the
    discretization error, which is why it is no longer decreasing. The uniform
    case contains $64^3$ cells, and the `center' and `corner' cases contain two
    additional levels of refinement, which if placed properly (the `center'
    case) indeed decreases the error.}
  \label{fig:conv-test}
\end{figure*}

Figure \ref{fig:conv-test} shows the residual $r = f - \nabla^2 \phi$ versus FMG
iteration. Two downward and two upward smoothing steps were taken per iteration.
The residual reduction factor per iteration is reduced when refinement is
present. This is a result of our ghost cell procedure near refinement
boundaries, see section \ref{sec:ghost-cells-boundary}, which reduces the
convergence rate. The residual is reduced up to machine precision after about 10
iterations. Due to the algorithmic steps involved, such as evaluating
expressions like equation \eqref{eq:poisson-2d}, the minimum residual that can
be obtained is proportional to $\epsilon_\mathrm{mach} h^{-2} |\phi|$, where
$\epsilon_\mathrm{mach} \approx 10^{-16}$ is the machine's precision, $h$ is the
mesh spacing and $|\phi|$ is the local amplitude of the computed solution.

Regardless of the lower reduction factor with refinement boundaries, figure
\ref{fig:conv-test} shows that the discretization error is reached in one or two
FMG iterations. The test problem has a steep Gaussian at the center, which is
where the largest discretization errors occur for the uniform grid case. With
the centered refinement errors are indeed significantly reduced. In the
$L_\infty$ norm, the error is reduced by a factor 20, slightly more than the
factor 16 expected from a second-order discretization with two levels of
refinement. With the corner refinement, the errors hardly change, showing that
`wrongly' placed refinements are handled well by our approach.

\subsection{Free space solutions in 3D}
\label{sec:free-space-solver}

\begin{figure}
  \centering
  \includegraphics[width=8cm]{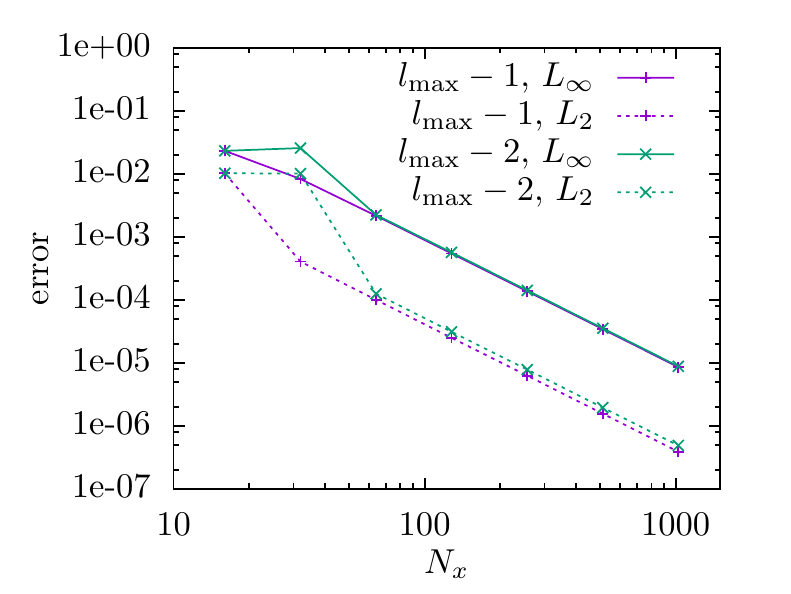}
  \caption{$L_2$ and $L_\infty$ norms of the error for the 3D free-space test
    problem given by equation \eqref{eq:gauss-rhs}. The error is shown versus
    the number of grid points per dimension. The labels $l_\mathrm{max}-1$ and
    $l_\mathrm{max}-2$ indicate at which grid level the direct free-space solver
    was called to obtain boundary conditions for the multigrid procedure, where
    $l_\mathrm{max}$ denotes the highest grid level.}
  \label{fig:free-space-error}
\end{figure}

To test our method with free space boundary conditions in 3D, see section
\ref{sec:free-space-boundary}, we solve a free space Poisson problem with a
Gaussian right-hand side
\begin{align}
  \label{eq:gauss-rhs}
  \nabla^2 \phi &= f,\\
  f(\vec{r}) &= \frac{-1}{\sigma^3 \pi^{3/2}} \exp(-|\vec{r}-\vec{r}_0|^2/\sigma^2),\nonumber
\end{align}
where $\vec{r}_0$ is the center of the Gaussian and $\sigma$ controls its width.
The solution is then given by
\begin{equation}
  \label{eq:gauss-sol}
  \phi(\vec{r}) = \frac{1}{4 \pi} \erf(|\vec{r}-\vec{r}_0|/\sigma) / |\vec{r}-\vec{r}_0|,
\end{equation}
where $\erf$ denotes the error function. The computational domain is again a
unit cube centered at the origin, and the Gaussian is located at the origin with
$\sigma = 0.1$. Figure \ref{fig:free-space-error} shows the $L_2$-norm of the
error after two FMG cycles versus the grid resolution. Two curves are shown, for
which the direct solver is called on levels $l_\mathrm{max}-1$ and
$l_\mathrm{max}-2$ respectively, where $l_\mathrm{max}$ denotes the highest grid
level. For grids larger than $64^3$, there is hardly any difference between the
two curves, and they show second order convergence. For simplicity, uniformly
refined grids are used, but the multigrid solver also works for AMR meshes.

The cost of the direct solver is relatively small, because it is only called
once per right-hand side on a coarser grid, and because the direct solver itself
is quite efficient~\cite{genovese_2006}.

\subsection{Strong scaling tests}
\label{sec:scaling-tests}

We now look at the performance and scaling of the geometric multigrid library,
by solving Poisson's equation with unit right-hand side
\begin{equation}
  \label{eq:poisson-scaling}
  \nabla^2 \phi = 1,
\end{equation}
on a unit cube, with $\phi$ set to zero at the boundaries. We measure the time
per multigrid cycle for both FMG and V-cycles by averaging over 100 cycles. For
both types of cycles two upward and two downward smoothing steps with a Jacobi
smoother were performed, and octree blocks of size $16^3$ were used. The scaling
results presented below were obtained on nodes with two 14-core Intel Xeon
E5-2680v4 processors, for a total of 28 cores per node. In all tests, one MPI
process per core was used. We present \emph{strong scaling} results, which means
that the problem size is kept fixed but the number of processors is increased.

Normally, the library copies its load balancing from another application, as
discussed in section \ref{sec:parallelization}. For the tests presented below,
the library was used by itself, in which case it performs a division of blocks
over processors similar to a Morton order, which is also used in
\texttt{MPI-AMRVAC}~\cite{morton_1966,keppens_2012}.

\begin{figure*}[ht]
  \centering
  \includegraphics[width=8cm]{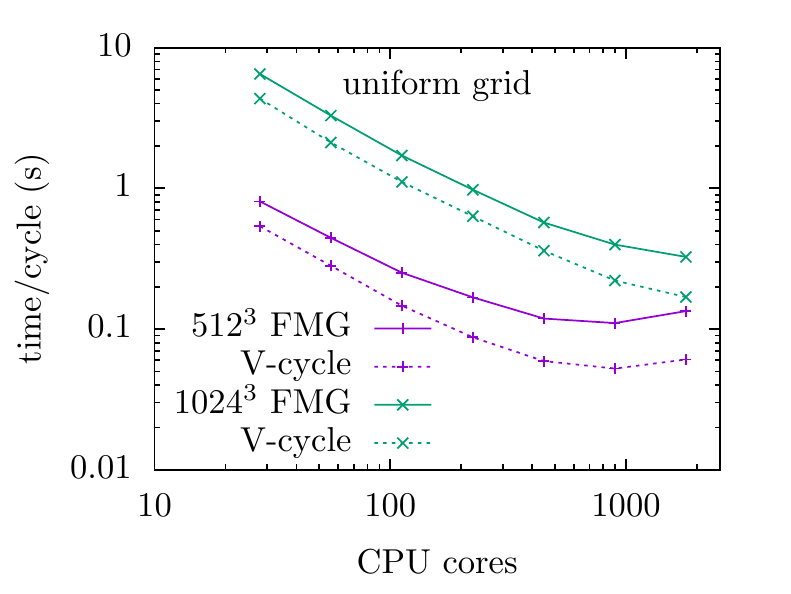}%
  \includegraphics[width=8cm]{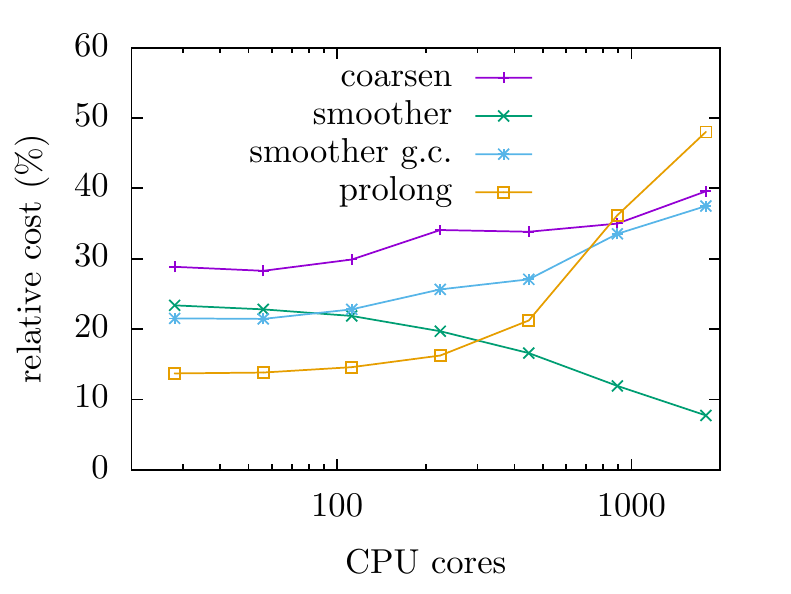}
  \caption{Left: strong scaling results for a problem size of $512^3$ and
    $1024^3$, showing the time per FMG cycle and per V-cycle. Right: breakdown
    of the computational cost of an FMG cycle for a $1024^3$ grid versus the
    number of CPU cores used.}
  \label{fig:strong-scaling-uniform}
\end{figure*}

Figure~\ref{fig:strong-scaling-uniform} shows strong scaling results for a uniform
grid containing either $512^3$ or $1024^3$ cells. For the $512^3$ case, scaling
results look good up to 448 cores, but with 1792 cores the performance is worse
than with 896 cores. This makes sense: even on the finest grid, the number of
unknowns is only about $42^3$ using 1792 cores. For the $1024^3$ test case,
scaling is closer to ideal (i.e., closer to a straight line in the figure), even
using 1792 cores. The figure also shows that the cost of a V-cycle is lower than
that of an FMG cycle. The difference increases with the number of cores, due to
the extra work on coarse grids with FMG cycles, see figure \ref{fig:mg-cycles}.

A breakdown of the relative cost of an FMG cycle for the $1024^3$ case is also
shown in figure \ref{fig:strong-scaling-uniform}. Shown is the percentage of
time spent on the transfer to coarse grids, the smoother (excluding
communication), the ghost cell exchange during smoothing steps (labeled smoother
g.c.), and the prolongation to finer grids. With an increasing number of cores,
less time is spent on computation compared to communication.

\begin{figure*}
  \centering
  \begin{minipage}{8cm}
    \includegraphics[width=\textwidth]{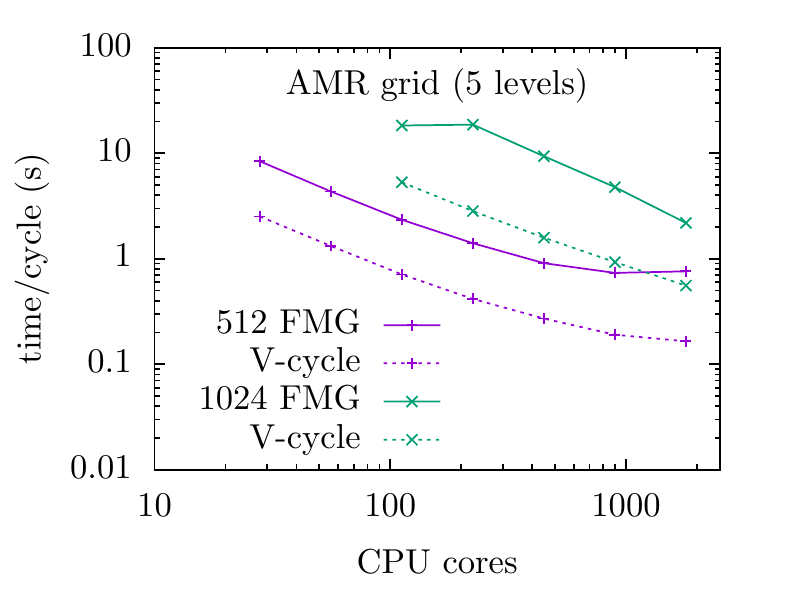}%
  \end{minipage}%
  \qquad%
  \begin{minipage}{4.8cm}
    \includegraphics[width=\textwidth]{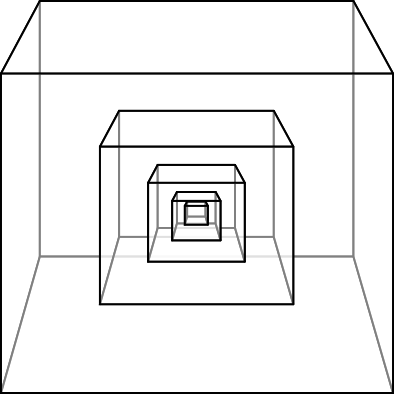}%
  \end{minipage}
  \caption{Strong scaling results on an AMR grid with 5 levels, with each
    level containing $512^3$ or $1024^3$ cells. The grid structure is illustrated
    on the right.}
  \label{fig:strong-scaling-refinement}
\end{figure*}

Figure \ref{fig:strong-scaling-refinement} shows strong scaling results on a
refined grid with a total of five levels. Each grid level contains either
$512^3$ or $1024^3$ cells, and the refinements are placed at the center of the
domain, as illustrated in the figure. Compared to the uniform grid cases, there
are about five times as many unknowns. The duration of V-cycles is indeed about
five times longer, although the parallel scaling is improved due to the larger
total number of unknowns. The difference in cost between V-cycles and FMG cycles
is larger than for the uniform grid case, due to the extra work the FMG cycles
perform on coarse grids, which now contain a significant number of unknowns.

\section{Divergence cleaning}
\label{sec:divergence-cleaning}

This section is specifically about divergence cleaning in magnetohydrodynamic
(MHD) simulations. For readers not interested in this particular application, we
provide a short summary of the main results below:
\begin{itemize}
  \item We compare elliptic, hyperbolic, and parabolic divergence cleaning for
  several test cases in 2D/3D Cartesian and $2.5$D cylindrical geometries, with
  AMR.
  \item With the multigrid library, divergence cleaning up to machine precision
  requires the magnetic field to be defined at cell faces. However, in
  \texttt{MPI-AMRVAC} the field is defined at cell centers, and we show that
  elliptic divergence cleaning can then still be applied successfully.
  \item We show that a fourth-order discretization of the right-hand side ($\nabla \cdot \vec{B}$) can be beneficial for elliptic divergence cleaning.
  \item The tests demonstrate the coupling of the multigrid solver to
  \texttt{MPI-AMRVAC}. The cost of runs was not significantly increased (by less
  than 10\%) when elliptic divergence cleaning was performed once per time step.
  \item The tests show that for typical problems, divergence cleaning methods
  play a similar role as the slope limiters: different methods give slightly
  different results and there is no single best method.
\end{itemize}

Maxwell's equations state that $\nabla \cdot \vec{B} = 0$, but this constraint
does not automatically hold in numerical MHD computations~\cite{Brackbill_1980}.
If no special care is taken, $\nabla \cdot \vec{B}$ can grow at each step
through discretization errors, leading to unphysical results. Therefore, a
number of methods has been developed to ensure $\nabla \cdot \vec{B}$ remains
small compared to discretization and truncation errors.

With the Hodge--Helmholtz projection method~\cite{Brackbill_1980,Marder_1987},
the divergence is cleaned by solving Poisson's equation:
\begin{align}
  \label{eq:poisson-divb}
  \nabla \cdot \nabla \phi = \nabla \cdot \vec{B}_\mathrm{old},\\
  \label{eq:poisson-divb-cleaning}
  \vec{B}_\mathrm{new} = \vec{B}_\mathrm{old} - \nabla \phi.
\end{align}
Below, we call this approach \emph{elliptic} divergence cleaning, and we will use multigrid to solve Poisson's equation. Another
approach is to add source terms to the MHD equations to control
$\nabla \cdot \vec{B}$ errors, as is done in the eight-wave formulation of
Powell~\cite{Powell_1999}, or its variants which only affect the induction equation~\cite{Janhunen_2000,Dellar_2001}. The MHD equations can also be modified to ensure transport and/or damping of $\nabla \cdot \vec{B}$ errors. The extra terms
can have a \emph{parabolic} (diffusive) character, as in the `diffusive' method
described in~\cite{Keppens_2003} which only adds a diffusion term to the induction equation. When using an extended version of the MHD equations with a variable that links to $\nabla \cdot \vec{B}$ error damping and transport, the method can also have a \emph{hyperbolic}
character, as in the Generalized Lagrange multiplier (GLM) method described
in~\cite{Dedner_2002} and the recently derived ideal GLM-MHD scheme presented
in~\cite{Derigs_2018}.

Constrained transport (CT) methods~\cite{Evans_1988,Balsara_1999,Ryu_1998} were
designed to preserve $\nabla \cdot \vec{B} = 0$ up to machine precision, typically by
defining the magnetic field at cell faces and the electric field at cell
corners. Variants that do not rely on a staggered representation of the magnetic field have been discussed in~\cite{Toth_2000} . CT methods have been made compatible with adaptive
mesh refinement~\cite{Balsara_2001,Fromang_2006,Cunningham_2009,Miniati_2011,Olivares_2018}, but their implementation is
non-trivial. Moreover, while CT methods ensure one particular discretization of the monopole constraint in machine precision, any other discretization will show truncation errors in places of large gradients, and especially at discontinuities. For mesh-free computations, \cite{Hopkins_2016} recently advocated the use of a constrained-gradient method, which in essence uses an iterative least-square minimization involving the magnetic field gradient tensor. Mesh-free smoothed-particle MHD implementations have also successfully devised constrained hyperbolic/parabolic divergence cleaning methods, where the wave cleaning speeds become space and time dependent~\cite{Tricco_2016}.

An extensive comparison of $\nabla \cdot \vec{B}$ cleaning techniques was
performed in~\cite{Toth_2000}, where a suite of rigorous test problems on uniform Cartesian grids showed that a projection scheme could rival central difference and constrained transport schemes in accuracy and reliability.  Further comparisons have been performed in
e.g.~\cite{Balsara_2004,Zhang_2016}, where especially~\cite{Balsara_2004} demonstrates some deficiencies in using divergence cleaning steps versus CT, when applied to supernova-induced MHD turbulence.

In the tests below, we compare \emph{elliptic}, \emph{parabolic} and
\emph{hyperbolic} divergence cleaning. `Elliptic' refers to the multigrid-based
projection method described in the next section, `parabolic' to the diffusive
approach of~\cite{Keppens_2003}, and `hyperbolic' to the EGLM-MHD method
described in~\cite{Dedner_2002}. For the EGLM-MHD approach, we set the parameter
$c_h$ to the globally fastest wave speed, and we use $c_p^2/c_h = 2h$ to
balance decay and transport of the $\psi$ variable, where $h$ is the finest grid
spacing. 

Below, a suffix 4\textsuperscript{th} indicates that $\nabla \cdot \vec{B}$
terms have been computed with a fourth-order accurate scheme, which is relevant
for the elliptic and parabolic methods.

\subsection{Elliptic divergence cleaning}
\label{sec:our-divb-method}

In \texttt{MPI-AMRVAC}, the magnetic field is defined at cell centers. To compute its
divergence in a Cartesian geometry, we consider two discretizations for
$\nabla \cdot \vec{B} = \partial_x B_x + \partial_y B_y + \partial_z B_z$. Each
derivative can either be approximated with second order central differences
\begin{equation}
  \label{eq:divb-central-diff}
  \partial_x B_x \approx \frac{B_{x,i+1} - B_{x,i-1}}{2 \Delta x},
\end{equation}
or with a fourth-order differencing scheme
\begin{equation}
  \label{eq:divb-4th-order}
  \partial_x B_x \approx \frac{-B_{x,i+2} + 8 B_{x,i+1}
    - 8 B_{x,i-1} + B_{x,i-2}}{12 \Delta x}.
\end{equation}
Afterwards, we update the magnetic field according to equation
\eqref{eq:poisson-divb-cleaning}, and update the energy density as
\begin{equation}
  \label{eq:divb-pressure-fix}
  e_\mathrm{new} = e_\mathrm{old} + \frac{1}{2}
  \left(B^2_\mathrm{new} - B^2_\mathrm{old}\right),
\end{equation}
which keeps the thermal pressure constant~\cite{Toth_2000}, which can be
important to avoid negative pressures. A downside is that equation
\eqref{eq:divb-pressure-fix} does not conserve total energy. For the correction
step, we evaluate $\nabla \phi$ with second-order central differences.

The multigrid solver described in this paper is cell-centered, and with its
standard 5/7-point stencil the divergence $\nabla \cdot \nabla \phi$ is computed
from a face-centered quantity ($\nabla \phi$). Since in \texttt{MPI-AMRVAC}
$\nabla \cdot \vec{B}$ is the divergence of a cell-centered quantity, the two
divergences do not exactly match\footnote{In principle, it is possible to use an
  operator with a wider stencil to ensure $\nabla \cdot \vec{B} = 0$ up to
  machine precision. However, this would make the solver more costly and also
  lead to a decoupling of unknowns, as discussed in e.g.~\cite{Balsara_2004}.}.
This means that after the projection step, $\nabla \cdot \vec{B}$ will be
non-zero in both the second and fourth order schemes, although differences will
be small for smooth profiles. Based on the results presented here and those
of~\cite{Toth_2000}, we think this is not necessarily a problem.

With a staggered discretization, in which the magnetic field is defined at cell
faces, the two divergences in equation \eqref{eq:poisson-divb} exactly match.
Divergence cleaning can then be performed up to machine precision. The multigrid
library is currently used in the \texttt{BHAC} code~\cite{Olivares_2018}, which
employs such a staggered discretization, to ensure that initial magnetic fields
are divergence-free up to machine precision.


\subsection{Field loop advection}
\label{sec:field-loop-advection}

We first consider the 2D field loop advection test described
in~\cite{Gardiner_2005}, which was also used more recently in
e.g.~\cite{Hopkins_2016}. A weak magnetic field loop is advected through a
periodic domain given by $x \in [-1,1]$ and $y \in [-1/2, 1/2]$. The initial
conditions are $\rho = 1$, $p = 1$, $(v_x, v_y) = (2, 1)$, and the magnetic
field is computed from a vector potential whose only non-zero component is
\begin{equation*}
A_z =
  \begin{cases}
     A_0 (R_0 - \sqrt{x^2 + y^2}) & \text{for } x^2 + y^2 \leq R_0^2\\
    0 & \text{for } x^2 + y^2 > R_0^2
  \end{cases},
\end{equation*}
where $A_0 = 10^{-3}$ and $R_0 = 0.3$. We numerically evaluate
$B_x = \partial_y A_z$ and $B_y = -\partial_x A_z$ using second-order central
differencing. Inside the magnetized field loop the plasma
beta is $\beta = 2 p/B^2 = 2 \cdot 10^6$, so that this is effectively a
hydrodynamics problem in which the magnetic field is a passive scalar.
Nevertheless, its solution can be sensitive to the divergence cleaning method
used~\cite{Gardiner_2005,Hopkins_2016}.

\begin{figure*}[ht]
  \centering
  \includegraphics[width=14cm]{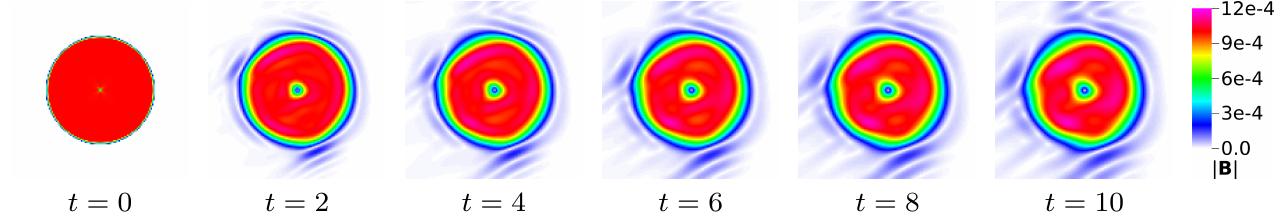}
  \caption{Example of the evolution of the magnetic field strength for the
    advected field loop test with parabolic divergence cleaning and the \v{C}ada
    slope limiter. At $t = 10$, the loop has translated 10 times (horizontally
    and vertically). The figures show half of the computational domain, namely
    $x \in [-1/2,1/2]$ and $y \in [-1/2,1/2]$.}
  \label{fig:field-loop-evolution}
\end{figure*}

\begin{figure*}[ht]
  \centering
  \includegraphics[width=12cm]{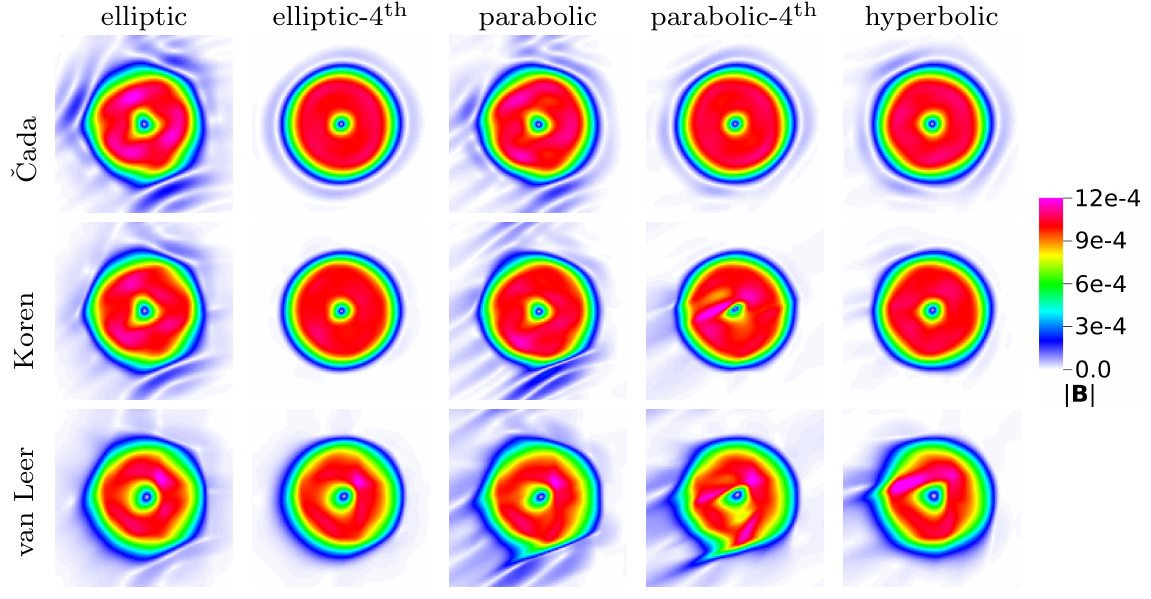}
  \caption{Magnetic field strength at $t = 10$ for the advected field loop test.
    The columns correspond to different divergence-cleaning methods, and a
    suffix 4\textsuperscript{th} indicates a fourth-order scheme is used to
    evaluate $\nabla \cdot \vec{B}$ terms. The rows correspond to different
    slope limiters. The figures show half of the computational domain, namely
    $x \in [-1/2,1/2]$ and $y \in [-1/2,1/2]$.}
  \label{fig:field-loop-bfield}
\end{figure*}

We simulate this system up to $t = 10$ on a uniform $256\times 128$ grid (AMR
tests follow in the next subsection), with a grid spacing $h = 1/128$. An
example of the evolution is shown in figure \ref{fig:field-loop-evolution},
which was obtained using the parabolic approach. Fluxes were computed with the
HLL scheme, using a CFL number of $0.5$. For figure
\ref{fig:field-loop-evolution}, a \v{C}ada limiter~\cite{Cada_2009} was used to
reconstruct cell face values for flux computations. At $t = 10$, the field loop
has moved through the system 10 times. We run simulations with several
combinations of slope limiters and $\nabla \cdot \vec{B}$ methods, always
employing the same HLL scheme. These slope limiters are used in \texttt{MPI-AMRVAC} to
reconstruct cell-face values from cell-centered ones for the flux
computation~\cite{keppens_2012}. Figure \ref{fig:field-loop-bfield} shows the
magnetic field strength $|\vec{B}|$ at $t = 10$ for three types of limiters,
described in~\cite{Cada_2009} (`\v{C}ada'), \cite{koren_limiter} (`Koren'), and
\cite{Van_Leer_1977} (`van Leer'), for five different cleaning approaches.

The computational cost of the \texttt{MPI-AMRVAC} runs hardly depended on the divergence
cleaning method that was used, with run times differing by less than 10\%. The
choice of limiter had a greater impact. Runs with the more complex `\v{C}ada'
limiter took up to 40\% longer than those with the simple van Leer limiter
(which is symmetric), and runs with the Koren limiter were in between.

\begin{figure*}
  \centering
  \includegraphics[width=8cm]{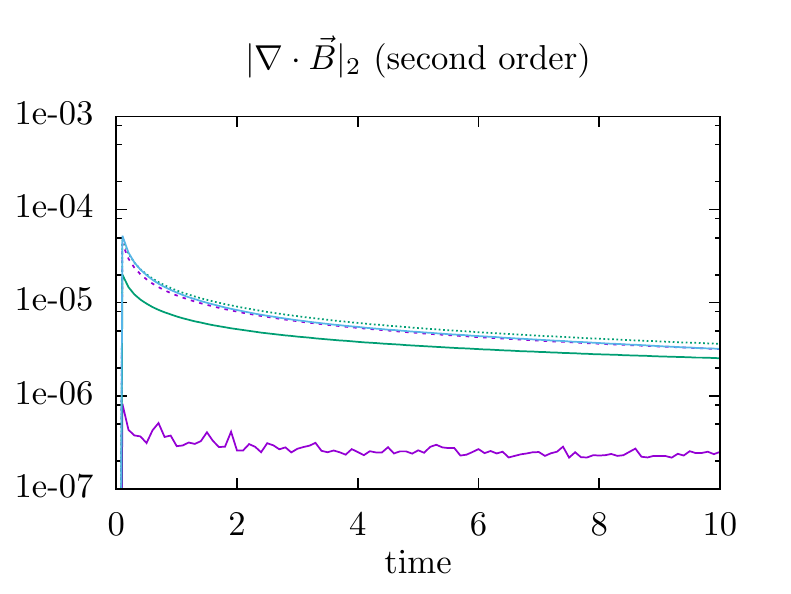}%
  \includegraphics[width=8cm]{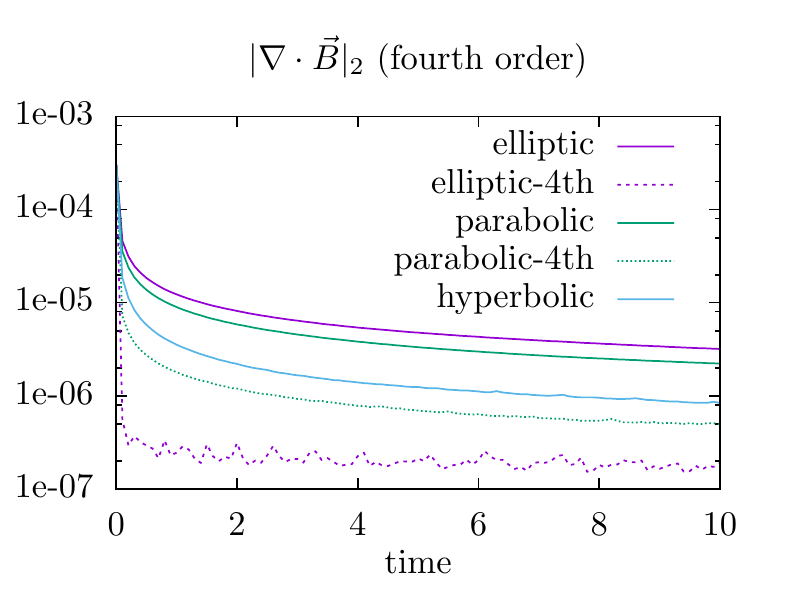}%
  \caption{The $L_2$-norm of $\nabla \cdot \vec{B}$ versus time for the advected
    field loop test on a uniform $256 \times 128$ grid. We only show the results
    for the \v{C}ada limiter. On the left, the $L_2$-norm is computed using a
    second order scheme for $\nabla \cdot \vec{B}$, on the right using a fourth
    order scheme.}
  \label{fig:field-loop-plots}
\end{figure*}

With a second-order evaluation of $\nabla \cdot \vec{B}$, elliptic and
hyperbolic divergence cleaning give similar results, somewhat worse than those
obtained with hyperbolic cleaning. With a fourth-order evaluation of
$\nabla \cdot \vec{B}$, elliptic cleaning gives the best results, which also
appear to be less sensitive to the limiter used. It is to be noted that more
structure is visible in Fig.~\ref{fig:field-loop-bfield} outside the loop than
shown in e.g.~\cite{Hopkins_2016}, but this is due to the combination of using a
different color scheme (our color legend is indicated in the figure), and
because we show $|\vec{B}|$ instead of $B^2$.


Figure \ref{fig:field-loop-plots} shows the $L_2$-norm of
$\nabla \cdot \vec{B}$, defined as
\begin{equation}
  \label{eq:l2-norm-example}
  |\nabla \cdot \vec{B}|_2 = \sqrt{\frac{1}{V} \int |\nabla \cdot \vec{B}|^2 dV},
\end{equation}
using a second-order and a fourth-order cell-centered evaluation. The elliptic
schemes give the smallest $|\nabla \cdot \vec{B}|_2$ when the same
second/fourth-order discretization is used to evaluate
$|\nabla \cdot \vec{B}|_2$ and the right-hand side of
Eq.~\eqref{eq:poisson-divb}. However, $\nabla \cdot \vec{B}$ being small in one
discretization does not mean it is small in another one, as was as observed in
\cite{Toth_2000}.

\begin{figure}
  \centering
  \includegraphics[width=8cm]{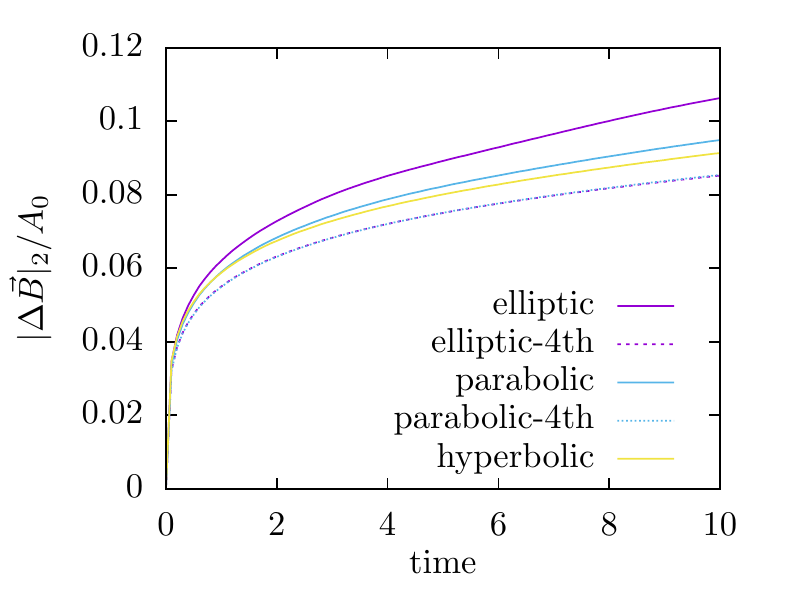}
  \caption{The $L_2$-norm of the error in the magnetic field
    $|\vec{B} - \vec{B}_\mathrm{sol}|$ divided by the initial amplitude
    $A_0 = 10^{-3}$, shown for the advected field loop test with the \v{C}ada
    limiter.}
  \label{fig:field-loop-Berror}
\end{figure}

The $L_2$-norm of $|\vec{B} - \vec{B}_\mathrm{sol}|$ is shown in figure
\ref{fig:field-loop-Berror}, where $\vec{B}_\mathrm{sol}$ is the approximate
solution to the problem, only taking into account advection of the initial
condition. From this comparison, the elliptic-4\textsuperscript{th} and
parabolic-4\textsuperscript{th} schemes give the best results, whereas the
standard elliptic and parabolic schemes perform a bit worse than other methods.
Since the cost of a fourth-order evaluation of $\nabla \cdot \vec{B}$ is
negligible, the results suggest that such an evaluation can be recommended. In
conclusion, we find that elliptic divergence cleaning works well to control
monopole errors for this test case.


\subsection{Advecting a current-carrying cylinder (mach 0.5)}
\label{sec:advect-curr-carry}

Although the previous setup became a popular test for magnetic divergence control, it is not a realistic test case for MHD applications since it only probes the very high plasma beta (order one million to infinity) regime. Moreover, the initial condition is not a true MHD equilibrium, has an infinite current at $r=0$ together with a current sheet at its boundary $R_0$, and the radially inward Lorentz force must actually set up sausage type compressions of the loop, which could well be responsible for the fluctuations seen in the environment of the advected loop in Fig.~\ref{fig:field-loop-evolution}.

Below, we introduce a more realistic advection test, which can be used to
demonstrate a number of typical computational challenges in MHD applications:
(1) combining high and low plasma beta regimes, (2) ensuring force balance, and
(3) handling surface current contributions in AMR evolutions.

\subsubsection{Description of the general test case}

We set up a current-carrying magnetic flux tube embedded in a uniform,
magnetized external medium, ensuring that a true MHD equilibrium is realized.
This is then further combined with a uniform flow field, that addresses whether
Galilean invariance is obtained.

Using the scale invariance of the MHD equations~\cite{BookHans}, we exploit units where the radius of the flux tube is equal to unity, where the density external to the loop is fixed at $\rho_{\mathrm{ext}}=1$, while the external plasma pressure is $p_{\mathrm{ext}}=1/\gamma$. This makes the external sound speed and its unit length crossing time the reference speed and time unit, respectively. The initial flow field is then controlled fully by its Mach number $M_0$ and orientation angles $\varphi_0$ and $\theta_0$, such that the constant speed components are found from
\begin{eqnarray}
v_x(t=0) &= &M_0 \sin \theta_0 \cos \varphi_0 \,,\\
v_y(t=0) &= &M_0 \sin \theta_0 \sin \varphi_0 \,,\\
v_z(t=0) &= &M_0 \cos \theta_0 \,.
\end{eqnarray}
We align the flux tube with the $z$-direction, and use a $[-L,L]^2$ fully periodic domain, where we resort to a 2.5D (invariance in $z$) computation, although the problem can also be simulated in full 3D.
The external medium has a uniform magnetization, which is determined by the corresponding inverse plasma beta parameter $\beta_{\mathrm{ext}}^{-1}$ as $B_{z,\mathrm{ext}}=\sqrt{2 \beta_{\mathrm{ext}}^{-1}/\gamma}$.

The flux tube itself has internal variation, and is a force-free cylindrical equilibrium introduced in~\cite{GoldHoyle_1960}, where the physics of solar flares was discussed. For solar coronal applications, ensuring a force-free equilibrium, which guarantees $\vec{J}\times\vec{B}=\vec{0}$ without enforcing a vanishing current $\vec{J}=\nabla\times\vec{B}$, is a typical computational challenge. The internal variation for $r=\sqrt{x^2+y^2}\le 1$ is given by
\begin{eqnarray}
\rho_{\mathrm{int}}(r) & = & \rho_0 (1-(1-d)r^2) \,, \\
B_{z,{\mathrm{int}}}(r) & = & \frac{B_0}{1+c^2 r^2} \,, \\
B_{\theta,{\mathrm{int}}}(r) & = & \frac{c r B_0}{1+c^2 r^2} \,.
\end{eqnarray}
The parameters $\rho_0$, $B_0$ and $c$ are best controlled by dimensionless numbers which quantify the pitch and strength of the magnetic field variation. The $d$ parameter quantifies the internal density contrast $d=\rho(1)/\rho(0)$. Introducing the $q$-factor at the tube radius $$q(1)=\frac{\pi B_z(1)}{L B_{\theta}(1)},$$ along with the plasma beta at the flux tube radius $\beta(1)$, as well as the ratio $R$ of the Alfv\'en speed at $r=0$ to the external sound speed, we can deduce that
\begin{eqnarray}
c & = & \frac{\pi}{L q(1)} \,, \\
p_{\mathrm{int}} & = & \frac{\beta(1)(1+\beta_{\mathrm{ext}}^{-1})}{\gamma(\beta(1)+1)} \,, \\
B_0 & = & \sqrt{\frac{2 p_{\mathrm{int}} (1+c^2)}{\beta(1)}} \,, \\
\rho_0 & = & \frac{B_0^2}{R^2} \,.
\end{eqnarray}
The flux tube is internally force-free and represents a nonlinear force-free field configuration where $\vec{J}=[2c/(1+c^2r^2)]\vec{B}$, while there is a constant pitch $q(r)=q(1)$. The embedded configuration is fully force-balanced since the above relations enforce the total pressure balance across the loop radius.

Any combination of input parameters $M_0$, $\theta_0$, $\varphi_0$,
$\beta_{\mathrm{ext}}^{-1}$, $q(1)$, $\beta(1)$, $d$, and $R$ represents a
meaningful test for which the exact solution is known: the flux tube will be
advected at the prescribed constant speed. These parameters could explore
regimes that are particularly challenging for numerical treatments, like taking
the pitch such that the flux tube is liable to kink instability, or advecting at
highly supersonic speeds, or verifying very low beta behavior, etc. The edge of
the flux tube carries a surface current, where density, pressure and magnetic
field components change discontinuously. This is typical for many solar,
astrophysical or laboratory plasma configurations.

\subsubsection{Results for a particular sets of parameters}

We here focus on the particular case where $L=2$, $d=0.05$, $M_0=0.5$ (i.e. Mach
0.5 advection), $\phi_0=45^\circ$, $\theta_0=70^\circ$, $\beta(1)=0.05$ (i.e. a
truly low beta flux tube), $q(1)=1.2$ (such that it is stable to external kink
modes through the Kruskal--Shafranov limit), $R=1$, and taking
$\beta_{\mathrm{ext}}^{-1}=0.05$ (i.e. a high beta surrounding medium). We
perform 2.5D simulations using a three-step Runge-Kutta integrator with the HLL
scheme combined with a Koren limiter, and a Courant parameter of $0.8$.
Parabolic and elliptic divergence cleaning is applied, using a fourth order
discretization of $\nabla \cdot \vec{B}$ terms. We use a base resolution per
direction of 128 with four grid levels, which effectively gives a $1024^2$
resolution. We run until normalized time $t=10$, at which time the flux tube is
almost advected back to its original position. Grid refinement is handled as
follows: we enforce the maximal refinement level to resolve the region that is
initially between $0.9<r<1.1$, to accurately treat the surface discontinuities
during the entire evolution.

\begin{figure}
\begin{center}
  \includegraphics[width=8cm]{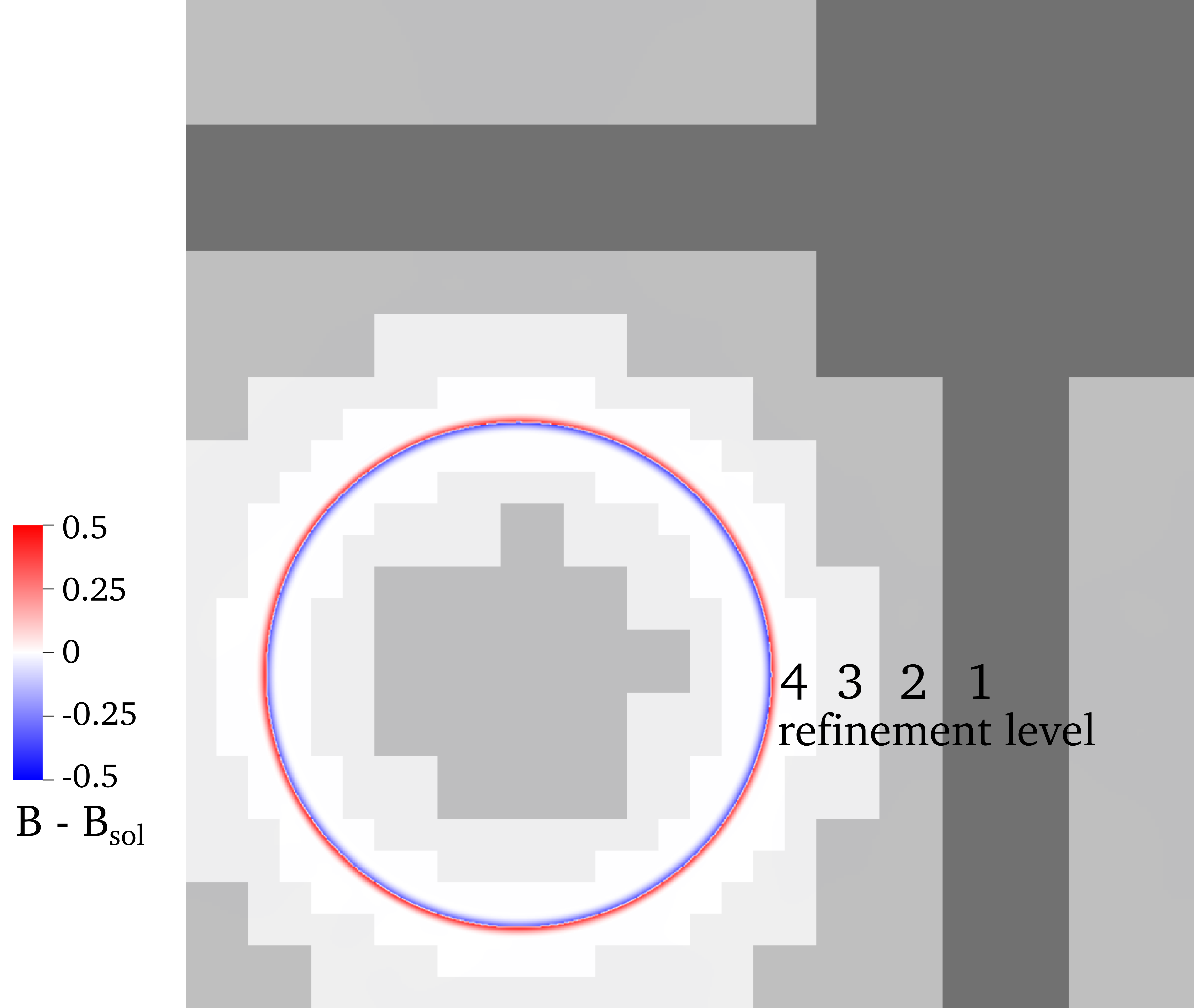}
  \caption{The error in the magnetic field magnitude at $t=10$ for the
    Gold--Hoyle force-free flux rope advection test, shown in the full $[-2,2]^2$
    domain. The errors are localized at the surface of the flux rope, where
    there is a surface current and a jump in density. The figure shows results
    for the multigrid approach (elliptic-4th). The four refinement levels are
    indicated by gray-to-white colors, with the finest (white) grid having a
    spacing of about $4 \times 10^{-3}$.}
 \label{fig:ccctest}
 \end{center}
\end{figure}

Figure~\ref{fig:ccctest} shows the error in the magnetic field strength at
$t = 10$ using elliptic divergence cleaning. The error is concentrated at the
boundary of the flux rope, where there is a surface current and a jump in
density. The grid structure at $t = 10$ is also shown in figure. With the
parabolic approach, the results are nearly identical.

\begin{figure}
  \begin{center}
    \includegraphics[width=\linewidth]{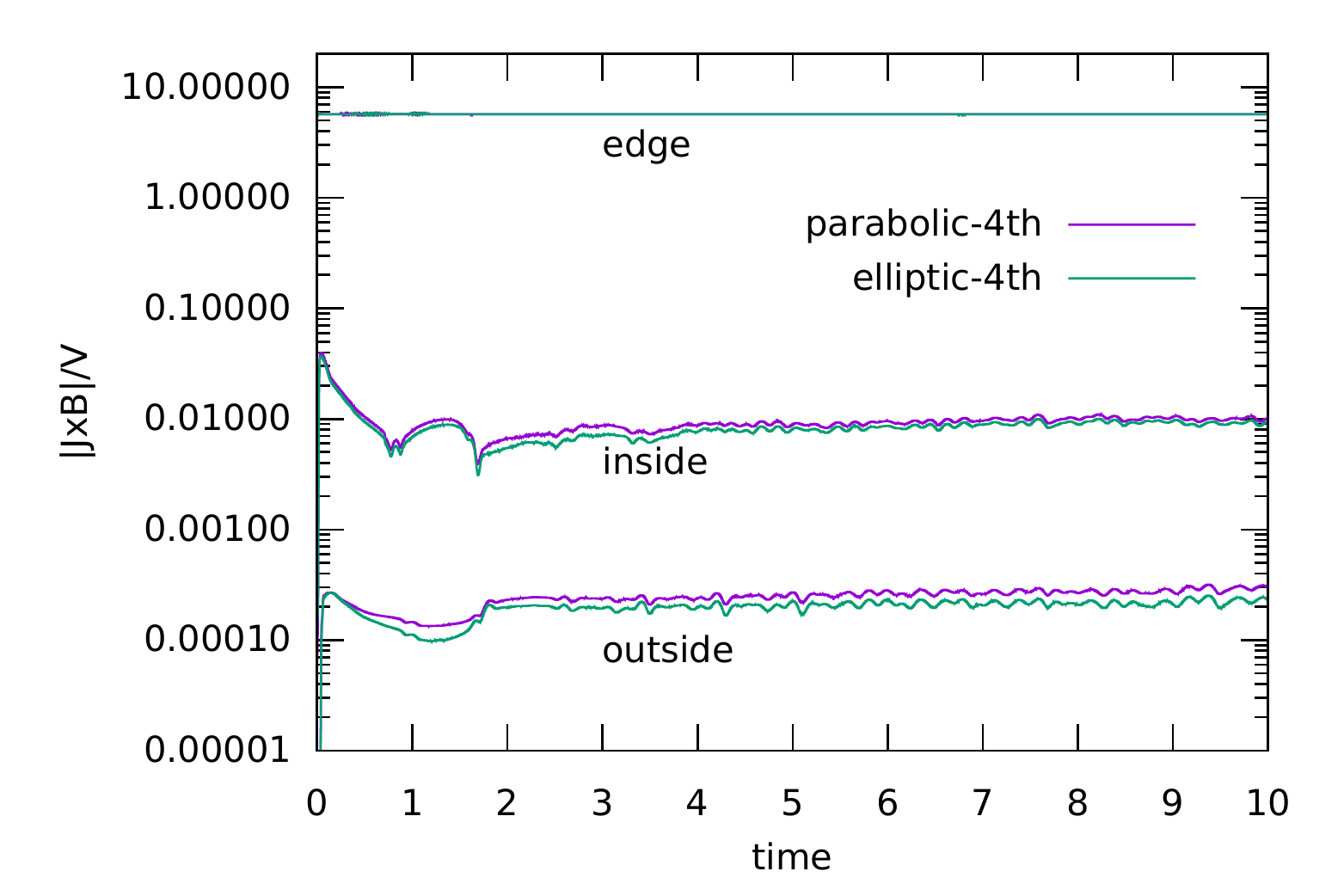}
    \caption{The average magnitude of $\vec{J} \times \vec{B}$ in three regions:
      the region outside the flux rope ($r > 1.1$), the region inside the flux
      rope ($r < 0.9$) and the region in between ($0.9 < r < 1.1$), where
      $r = 0$ corresponds to the center of the flux rope.}
    \label{fig:ccctest-regions}
  \end{center}
\end{figure}

Figure~\ref{fig:ccctest-regions} shows the average magnitude of
$\vec{J} \times \vec{B}$ in the region inside, outside and at the boundary of
the flux rope. Results are shown for both the elliptic and parabolic approach,
but only small differences between the two methods can be observed. Note that
the simulation is nearly force-free inside and outside the flux rope. At the
edge of the flux rope $\vec{J} \times \vec{B}$ is significantly larger, due to
the numerical discretization errors at the flux rope boundary.

We remark that with hyperbolic divergence cleaning, we obtain nearly identical
results. In conclusion, this test case shows that for a physically realistic
test case, the type of divergence cleaning has less effect than for the test
problem of section~\ref{sec:field-loop-advection}. It also demonstrates that our
divergence cleaning methods, and more specifically the elliptic approach, can
handle adaptive mesh refinement.

\subsection{Modeling magnetized jets (2.5D and 3D)}

A final demonstration of the multigrid-based divergence cleaning methodology
focuses on a typical astrophysical application: the propagation of a strongly
magnetized, supersonic and super-Alfv\'enic jet. We perform both 2.5D
axisymmetric and 3D Cartesian simulations. The configuration is borrowed
from~\cite{Keppens_2008}, where relativistic jets with helical field topologies
were studied in axisymmetry. We here use the same magnetic field topology in the
initial conditions, and evolve purely Newtonian cases, so we take parameters
similar to the non-relativistic jet case from~\cite{Keppens_2008}.

\subsubsection{Description of the test case}

At $t=0$, the jet occupies a finite region $R<R_j$ and $Z<Z_j$ where the (normalized) density is $\rho_j=1$, while the surrounding medium has a higher density $\rho_c=10$: the under-dense jet is entering a denser `cloud' region.
We take the domain size as follows: $R\in[0,30]$ and $Z\in[0,90]$, while $R_j=1.5$ and $Z_j=3.0$. The jet itself has a twisted field topology with an azimuthal field component in the jet region given by
\begin{equation}
B_\varphi=\tanh \frac{R}{5} \,,
\end{equation}
while this azimuthal field vanishes in the surroundings. Note again how this setup thereby necessarily involves surface current distributions (located at the edge of the jet region). The other magnetic field components are
\begin{eqnarray}
B_R & = & 2\frac{R_j}{Z_j} \frac{\left(\frac{Z}{Z_j}\right)^3 \tanh\left(\frac{Z}{Z_j}\right)^4\tanh\left(\frac{R}{R_j}\right)^2}{\frac{R}{R_j} \cosh\left(\frac{Z}{Z_j}\right)^4} \,, \\
B_Z & = & B_c + \frac{1}{\left[\cosh\left(\frac{R}{R_j}\right)^2\right]^2\cosh\left(\frac{Z}{Z_j}\right)^4} \,.
\end{eqnarray}
This magnetic field setup is analytically divergence-free (as it should), and ensures that the jet is entering an almost uniformly magnetized cloud region where the initial field strength has the `cloud' value $B_c=0.01$. The initial pressure distribution follows from
\begin{equation}
p= p_j + \frac{1}{2}-\frac{1}{2}\left(B_{\varphi}^2(R,Z)+B_Z^2(R,Z)\right) \,,
\end{equation}
where we take the jet pressure parameter $p_j=2$: this makes the internal jet region slightly under-pressured with respect to the external medium, and an order of magnitude hotter than its surroundings. Finally, the flow field $\vec{v}$ vanishes at $t=0$ outside the jet region, but within the jet follows from
\begin{eqnarray}
v_R & = & 0 \,, \\
v_Z & = & \alpha \frac{B_\varphi}{(R/5)\sqrt{\rho_j}} \,, \\
v_\varphi & = & \frac{B_\varphi}{\sqrt{\rho_j}} \,.
\end{eqnarray}
The parameter $\alpha=6.0$. These choices turn the jet Mach number $v_Z/c_s\approx 3$ while its Alfv\'en Mach number $v_Z/v_A\approx 6$ (both of these quantities vary with radius and relate to the local sound speed $c_s=\sqrt{\gamma p/\rho}$ and Alfv\'en speed $v_A=B/\sqrt{\rho}$). The ratio of specific heats is fixed at $\gamma=5/3$. In accord with the frequently invoked equipartition argument for astrophysical jets, the plasma beta internal to the jet is of order $\beta=2p/B^2 \approx 4$, while it is about 50000 in the cloud region.

\subsubsection{Computational domain and refinement}

In 2.5D, the resolution uses a coarse $32\times 64$ base grid, allowing a total of 6 AMR levels (i.e. an effective resolution of $1024\times 2048$). Refinement uses the Lohner estimator, this time taking in weighted information from $\rho$, $m_R=\rho v_R$ and $B_\varphi$ in a $0.5-0.25-0.25$ ratio. Maximal resolution is enforced within the region $R<3R_j$ and $Z<3Z_j$. Boundary conditions use the usual (a)symmetric combinations to handle the $R=0$ symmetry axis and extrapolate all variables at side and top in a zero-gradient fashion. The bottom boundary uses the analytic initial conditions within $R<R_j$, and adopts a reflective boundary beyond.

In 3D, our setup adopts the same physical parameters, but this time in a 3D Cartesian $(x,y,z)$ box of size $[-30,30]\times[-30,30]\times[0,90]$, where the $z$ axis coincides with the symmetry axis employed in the 2.5D runs (making $R=\sqrt{x^2+y^2}$). To avoid an artificial $m=2$ selection effect in the way non-axisymmetric modes with azimuthal mode number $m\ne 0$ develop from the noise (inherent to doing cylindrical problems on a Cartesian grid), we used a deterministic incompressible velocity perturbation consisting of 7 mode numbers $m=1,\ldots,7$ derived from $\psi=\sum_m A_m \cos(m\varphi+\phi_m)\exp(-[(R-0.75R_j)/R_j]^2)$ such that $\delta\vec{v}=\nabla\times \psi(x,y)\hat{e}_z$. This is applied in the ghost cells at the bottom ($z=0$) boundary only, where we add it to the fixed velocity field providing the jet conditions. The 7 amplitudes are chosen such that a maximal amplitude for each mode is $A_m\le 0.05$. In 3D, our base resolution is $64\times 64\times 96$, with 5 refinement levels to get to $1024\times 1024\times 1536$ effectively. Refinement in 3D is based on density only, augmented with user-enforced geometric criteria, where e.g.~the maximal resolution is always attained within the region $R<3R_j$ and $Z<3Z_j$. Boundary conditions at all sides and top extrapolate primitive variables using Neumann zero-gradient prescriptions. The bottom boundary fixes the entire initial condition, augmented with the $\delta\vec{v}$ addition, within the jet zone, while reflective boundaries are used beyond $R>R_j$.

\subsubsection{Results}

We run till time $t=60$, such that the jet progressed up to about $z\approx 60$.
We use a strong-stability preserving Runge-Kutta scheme (its implementation in
\texttt{MPI-AMRVAC} is demonstrated in~\cite{porth_2014}), an HLLC discretization, and
Piecewise Parabolic (PPM) reconstruction. Runs differ only in their divergence
cleaning approach. We anticipate many turbulent features related to fluid
instabilities, waves, rarefactions and shocks, as typical for under-dense
supersonic jets, but all differences in the jet morphology here entirely relate
to the error control on magnetic monopoles.

\begin{figure}
\begin{center}
\includegraphics[width=\linewidth]{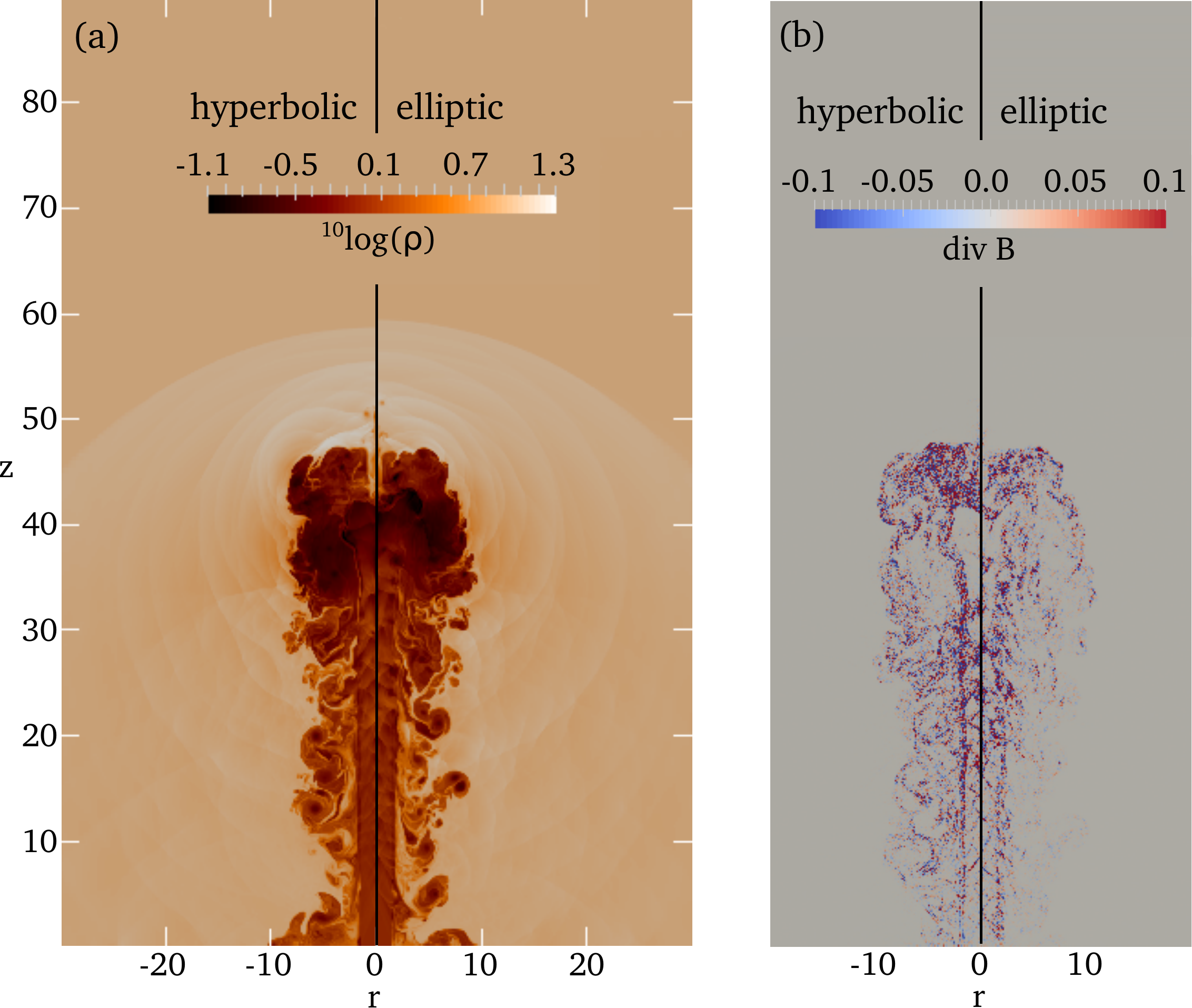}
\caption{(a) The logarithm of the density at $t=60$ for the helically magnetized
  jet in axisymmetric conditions, using the hyperbolic (left) versus the
  multigrid-based elliptic (right) treatment for monopole control. (b) The
  numerical value of $\nabla \cdot \vec{B}$, evaluated with a second order
  central difference formula, at $t=60$ for the hyperbolic and elliptic
  approach.}
\label{fig:jet-2d-plot}
\end{center}
\end{figure}

In Fig.~\ref{fig:jet-2d-plot}(a), we show the density distribution for the
axisymmetric simulations at $t = 60$, comparing the hyperbolic with the elliptic
method for divergence control. Naturally, many details differ between the two
cases, although both recover the richness in internal jet beam shocks, fluid
instabilities developing at the leading contact interface between jet and
surroundings, and the turbulent backflows where many vortical structures exist.
The shocked cloud matter is riddled with shocks. Repeated deformations of the
contact interface shed plasma into the backflow surrounding the jet spine.

A direct comparison of the monopole errors at $t = 60$ is given in
Fig.~\ref{fig:jet-2d-plot}(b). We here show a second order central difference
evaluation of $\nabla \cdot \vec{B}$ (we also used the second order evaluation
of the source term in the cleaning methods). With the elliptic cleaning there
are fewer cells with significant $\nabla \cdot \vec{B}$ values. All monopole
errors concentrate near the many discontinuities, as expected. Overall, the jet
progressed to about the same distance.

\begin{figure*}
  \centering
\includegraphics[width=14cm]{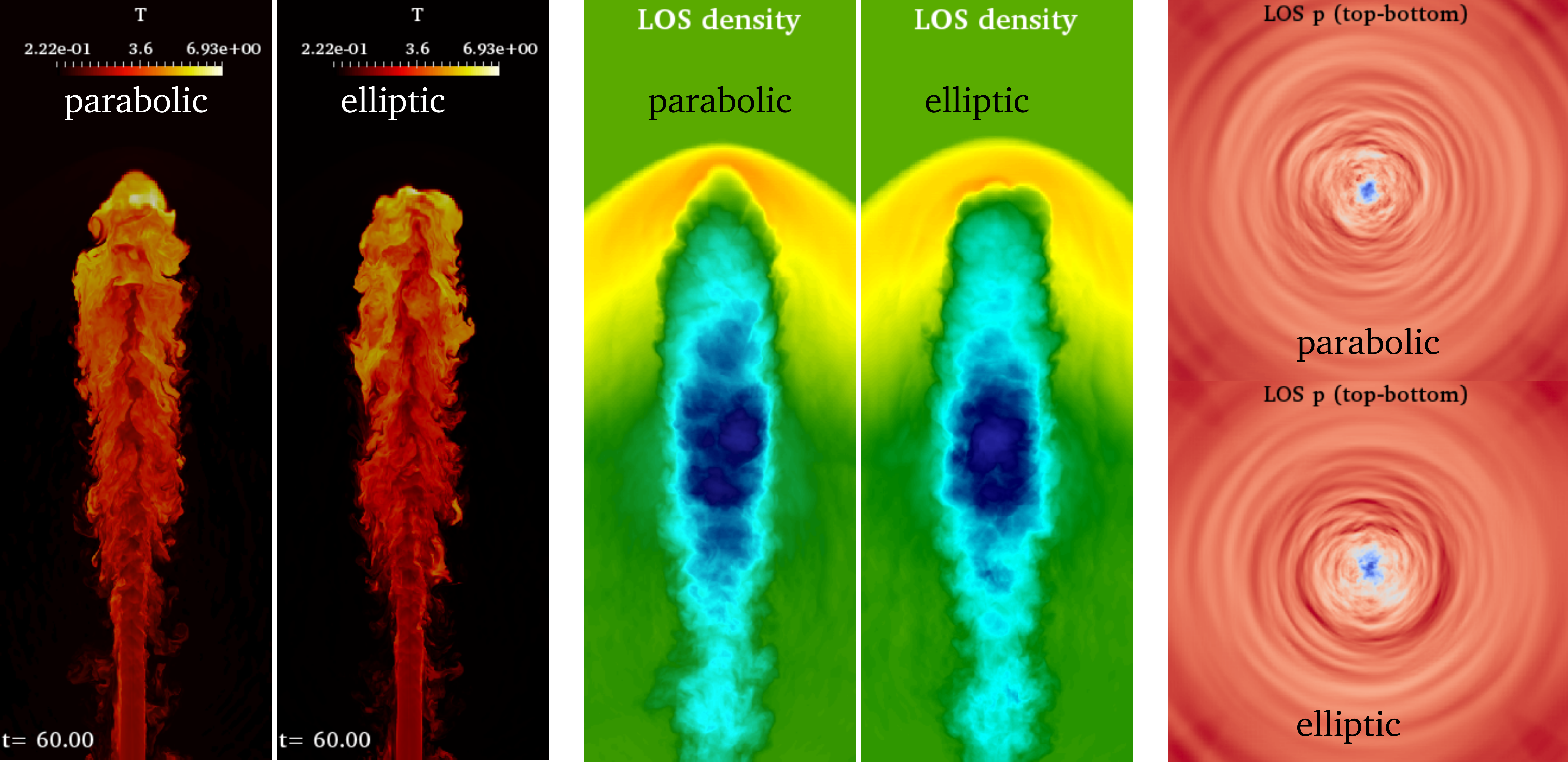}
\caption{Several views on the 3D jet simulation at $t=60$, where we used the
  parabolic and elliptic approach for divergence control. We show a
  cross-section of the temperature (left), a line-integrated side view of the
  density (middle) and a line-integrated top view of the pressure (right).}
  \label{fig:jet-3d-linde}
\end{figure*}

The same simulation in full 3D allows for non-axisymmetric deformations, which
can come about from current-driven kink instabilities mediated by the helical
magnetic field of the jet, or by the many shear-flow driven events. The state at
$t=60$ is shown in Fig.~\ref{fig:jet-3d-linde}, where we now compare the
elliptic approach to the parabolic one. The cross-sectional temperature view (left), and the line-of-sight integrated density views (middle) cover the full extent in $z\in[0,90]$, while the integrated pressure view shows the entire $x-y$ cross-section $[-30,30]^2$. The turbulent cocoon that develops
around the jet spine aids in retaining a coherent jet over the distance
simulated: the turbulence in the backflow region seems to prevent large
deformations of the jet. The overall morphology of the 3D helical jet is very
similar with both monopole corrections. A more in-depth discussion of the
physics in the context of astrophysical jet propagation is deferred to future
work. Fig.~\ref{fig:jet-3d-linde} shows that the temperature, density, and
pressure variations are all very well recovered with either method for monopole
control.

\section{Conclusions}

We have presented an MPI-parallel geometric multigrid library. The library can
be used to extend octree-based adaptive mesh refinement frameworks with an
elliptic solver. The library supports multigrid V-cycles and FMG cycles, and
employs standard second-order discretizations. Cartesian 2D/3D and cylindrical
2D grid geometries can be used, with periodic, Dirichlet, or Neumann boundary
conditions. For 3D Poisson problems free-space boundary conditions are also
supported, by using an FFT-based solver on the coarse grid. The convergence and
scaling of the library has been demonstrated with multiple test problems.

We have demonstrated the coupling of the library to \texttt{MPI-AMRVAC}, an
existing AMR code, by using the multigrid routines for divergence cleaning in
MHD simulations. We have compared three approaches: elliptic, hyperbolic and
parabolic divergence cleaning. Several test cases were presented, in 2D and 3D
Cartesian as well as axisymmetric geometries. Elliptic divergence cleaning
(i.e., using a projection method) was found to work satisfactorily in all cases,
although the other methods generally performed similarly well.

\paragraph{Acknowledgments}
JT is supported by postdoctoral fellowship 12Q6117N from Research Foundation --
Flanders (FWO). RK acknowledges support by FWO-NSFC grant G0E9619N.

The computational resources and services used in this work were provided by the
VSC (Flemish Supercomputer Center), funded by the Research Foundation --
Flanders (FWO) and the Flemish Government -- department EWI.

\bibliographystyle{elsarticle-num}
\bibliography{big_jannis}

\end{document}